\documentclass[11pt,dvipsnames]{article}
\usepackage{caption}
\usepackage{scalerel}
\usepackage{amsthm,latexsym,amsxtra,graphicx,appendix,epstopdf,feynmf,hyperref,setspace,fix-cm}
\usepackage[normalem]{ulem}
\usepackage{makeidx}
\usepackage{fullpage}
\usepackage{amsmath}
\usepackage{amssymb}
\usepackage{setspace}
\usepackage{bbm}
\usepackage{dsfont}
\usepackage{epsfig}
\usepackage[font=footnotesize,labelfont=bf,justification=centerlast,width=.94\textwidth]{caption}
\usepackage{cite}
 \usepackage{multirow}
\usepackage{array,booktabs,tabularx}
\usepackage{subcaption}
\usepackage{mathtools}
\usepackage{color}
\usepackage[makeroom]{cancel}
\usepackage{tikz}
\usepackage{pgfplots}
\pgfplotsset{compat=1.18}
\usepackage{bm}

\usepackage{tensor}
\usepackage{simplewick}
\usepackage{frcursive}

\usepackage{pgfornament,multicol}

\usepackage{caption}
\usepackage{hyperref}

\usepackage[skip=0.2cm, indent=0pt]{parskip}

\hypersetup{
    colorlinks,%
    citecolor=blue,%
    filecolor=blue,%
    linkcolor=blue,%
    urlcolor=blue
}
\usepackage{float}
\usepackage{slashed}
\usepackage{tikz}
\usetikzlibrary{decorations.pathmorphing}

\hypersetup{bookmarksdepth=subsection}
\usepackage{etoolbox}
\pretocmd{\appendix}{
  \addtocontents{toc}{\protect\setcounter{tocdepth}{1}}
  \hypersetup{bookmarksdepth=section}
}{}{}

\usepackage{tcolorbox}
\usepackage{tabularx}
\usepackage{array}
\usepackage{colortbl}
\tcbuselibrary{skins}

\tcbset{tab2/.style={enhanced,fonttitle=\bfseries,fontupper=\normalsize\sffamily,
colback=Goldenrod!20!white,colframe=Blue!50!black,colbacktitle=Goldenrod!50!white,
coltitle=black,center title}}

\def\r{\rightarrow}

\newcommand{\bi}{\begin{itemize}}
\newcommand{\ei}{\end{itemize}}
\newcommand{\bea}{\begin{eqnarray}}
\newcommand{\eea}{\end{eqnarray}}

\def\={\, = \,}

\def\frakr{\mathfrak{r}}

\def\frakt{\mathfrak{t}}
\def\bomega{\bm{\omega}}
\def\r{\mathfrak{r}}

\def\L{\mathfrak{L}}

\def\r{\mathfrak{r}}
\def\frakr{\mathfrak{r}}
\def\bomega{{\boldsymbol{\omega}}}

\def\XXint#1#2#3{{\setbox0=\hbox{$#1{#2#3}{\int}$}
     \vcenter{\hbox{$#2#3$}}\kern-.5\wd0}}

\def\={\, = \,}

\newcommand{\beq}{\begin{equation}}

\newcommand{\eeq}{\end{equation}}

\usepackage{tikz}
\usetikzlibrary{positioning}
\usetikzlibrary{intersections}
\usetikzlibrary{fadings} 
\usetikzlibrary{arrows.meta} 
\usetikzlibrary{arrows}

\tikzfading[name=fade out,
inner color=transparent!0,
outer color=transparent!100]

\definecolor{cherryblossompink}{rgb}{1.0, 0.72, 0.77}
\definecolor{lightblue}{rgb}{0.68, 0.85, 0.9}

\usetikzlibrary{decorations.pathmorphing}
\usetikzlibrary{decorations.pathreplacing,decorations.markings}

\usetikzlibrary{backgrounds,automata}

\newsavebox\CBox
\newcommand\hcancel[2][0.5pt]{%
  \ifmmode\sbox\CBox{$#2$}\else\sbox\CBox{#2}\fi%
  \makebox[0pt][l]{\usebox\CBox}%
  \rule[0.75\ht\CBox-#1/2]{\wd\CBox}{#1}}

\numberwithin{equation}{section}
\allowdisplaybreaks  %

\begin{document}

\vspace*{2.5cm}
\begin{center}
{ \huge {Umbilic Observatories} }\\

\vspace*{1cm}
\end{center}

\renewcommand{\thefootnote}{\fnsymbol{footnote}}

\begin{center}
Dionysios Anninos$^{1,2}$, Ra\'ul Arias$^3$, and Dami\'an A. Galante$^1$
\end{center}
\begin{center}
{
\footnotesize
\vspace{0.2cm}
$^1$Department of Mathematics, King's College London, Strand, London WC2R 2LS, UK
\\
$^2$Instituut voor Theoretische Fysica, KU Leuven, Celestijnenlaan 200D, B-3001 Leuven, Belgium
\\ 
$^3$Instituto de F\'isica La Plata -- CONICET and Departamento de F\'isica, \vspace{-1mm}
\\
Universidad Nacional de La Plata, C.C. 67, 1900 La Plata, Argentina
} 
\end{center}
\begin{center}
{\textsf{\footnotesize{
dionysios.anninos@kcl.ac.uk, rarias@fisica.unlp.edu.ar, damian.galante@kcl.ac.uk}} } 
\end{center}

\renewcommand{\thefootnote}{\arabic{footnote}}

\vspace*{0.5cm}

\vspace*{1.5cm}
\begin{abstract}
\noindent We study gravitational dynamics for the simplest example of an umbilic boundary in four-dimensional general relativity --- a three-dimensional de Sitter hypersurface embedded in a four-dimensional Minkowski world. Generally speaking, an umbilic surface has the property that its induced metric is proportional to its extrinsic curvature. We consider the case of both conformal boundary conditions, as well as Dirichlet boundary conditions imposed at the timelike hypersurface. For both cases, we find that there are no linearised modes whose imaginary frequency grows with angular momentum. The conformal boundary conditions allow for a dynamical boundary degree of freedom, in addition to the bulk excitations.  The umbilic property linearly decouples the boundary Weyl mode from the bulk dynamics, yielding linearly stable dynamics.  In contrast, the Dirichlet problem has no boundary mode. The umbilic hypersurfaces may thus constitute an improved starting point for defining gravitational dynamics in the presence of a timelike boundary.  Throughout our analysis, we draw parallels to the structure of spatial infinity for asymptotically Minkowski spacetimes and a more finite realisation thereof.  
\newline\newline 

\end{abstract}

\newpage

\setcounter{tocdepth}{2}
\tableofcontents

\section{Introduction} \label{sec:intro} 

The purpose of this article is to discuss the dynamics of four-dimensional gravitational theories with $\Lambda=0$ in the presence of a finite-size timelike boundary $\Gamma$. This is an open mathematical problem that has received considerable attention in recent years from both the mathematics \cite{Friedrich:1998xt,Anderson:2006lqb,Fournodavlos:2020wde, An:2021fcq, Fournodavlos:2021eye,An:2025gvr,An:2025cbs, An:2025rlw,Liu:2025xij} and physics \cite{Bredberg:2011xw, Anninos:2011zn,Coleman:2021nor,Silverstein:2024xnr,Philcox:2025faf,Anninos:2022ujl,Anninos:2023epi, Anninos:2024wpy, Anninos:2024xhc, Banihashemi:2024yye,Banihashemi:2025qqi, Galante:2025tnt, Allameh:2025gsa, Galante:2025emz, Anninos:2025zgr, Hamdan:2026olq,Callebaut:2025zpc} perspective. Here, we will consider gravitational perturbations in a four-dimensional Minkowski space  equipped with a three-dimensional de Sitter (dS$_3$) timelike boundary. A dS$_3$ boundary naturally resides near the spatial infinity of asymptotically Minkowski space \cite{AshtekarHansen1978,Sommers1978,Beig:1982ifu,ashtekar1992spatial,Compere:2011db,Henneaux:2018cst,Prabhu:2019daz,Capone:2022gme,Ashtekar:2023wfn}. When filled with the empty vacuum, the induced metric of dS$_3$  has the additional property of being proportional to the extrinsic curvature, a property known as umbilicity. Such timelike umbilic boundaries have the further interesting geometric feature that they are totally geodesic with respect to null rays, with the photon sphere being a prime example \cite{Cederbaum:2019rbv}.\footnote{From a different perspective, if the four-dimensional Minkowski world is a false vacuum subject to a Coleman--De Luccia nucleation event \cite{Coleman:1980aw}, and working in the thin wall regime, it is naturally equipped with a dS$_3$ timelike boundary. In a separate context, certain $\mathbb{Z}_2$ symmetric brane-world scenarios \cite{Randall:1999vf,Shiromizu:1999wj}, and end of the world tensionful branes \cite{Karch:2000ct}, also involve umbilic boundaries.}

In this article, we will predominantly consider the case of conformal boundary conditions, whereby the trace of the extrinsic curvature, $K$, is fixed alongside the conformal class of the boundary metric $[g]_{mn}$ and a suitable set of Cauchy initial data \cite{An:2021fcq}. One of the most striking features uncovered in the linearised analysis of gravitational perturbations around a Minkowski filling of a static boundary with round two-sphere cross-sections is the appearance of certain exponentially growing modes. Concretely, upon decomposing linearised gravitational perturbations into spherical harmonics labelled by a total angular momentum $l$, it was observed in \cite{Anninos:2023epi,Liu:2024ymn} that the frequency of perturbations (in units of the radius of the sphere) for $l\gg 1$ scales as  $\omega_l \approx l \pm i c l^{\frac{1}{3}}$ where $c$ is a real number. The growing modes have the additional feature that they are exponentially localised near the boundary of the wall, with a radial decay width that scales as $l^{-\frac{2}{3}}$. The growth is sufficiently fast, that their presence puts to question, in the sense of Hadamard \cite{hadamard1902problemes},  the mathematical well-posedness of the underlying partial differential equations governing the non-linear initial boundary value problem. The growing modes were further shown to be universally present for a large class of boundaries \cite{Anninos:2025zgr}. Nonetheless, placing the timelike boundary near a black hole/cosmological horizon or the boundary of AdS$_4$ has the notable feature of creating a hierarchic gap \cite{Anninos:2024xhc,Anninos:2025zgr} below which they cease to exist. In such settings, one can drive the minimal $l$ to be parametrically large before the onset of the aforementioned exponential growth, inviting a Wilsonian perspective on Hadamard's paradigm.

In any event, it is of interest to identify instances where the growing modes are absent altogether, at least at the linearised level. There are a few hints suggesting a scenario where this might happen. Firstly, in the analysis of \cite{An:2021fcq,Anninos:2023epi} it was observed that for a flat Lorentzian wall with a Minkowski filling no growing modes were present. Similarly, for a planar slicing of AdS$_4$, again the growing modes were absent \cite{Anninos:2024xhc}. Relatedly, in the strict global AdS$_4$ boundary one retrieves the ordinary stable AdS$_4$ problem \cite{Anninos:2024xhc}. Finally, in equation (B.8) of \cite{Anninos:2025zgr} it was observed that when the effective radial potential governing the linearised perturbations exhibits a critical point, as is the case for the photon sphere of a Schwarzschild black hole, the coefficient $c$ of the imaginary part in the frequency $\omega_l$ vanishes.\footnote{Linearised gravitational perturbations around the photon sphere of Schwarzschild black holes will be analysed elsewhere \cite{dipierro}.} All these examples have the geometric commonality that the timelike boundary is equipped with an induced metric that is of the umbilic type --- its induced metric is proportional to the extrinsic curvature of the infilling spacetime. A study of general relativity with boundary conditions enforcing the umbilic condition on all configurations was undertaken in \cite{Fournodavlos:2021eye}, for which the well-posedness of the initial boundary value problem was established. Here, instead, we will study perturbations around an umbilic boundary in a theory subject to either conformal or Dirichlet boundary conditions.

In what follows, we will explore the simplest example of a three-dimensional umbilic boundary. This is a dS$_3$ slicing of an ambient four-dimensional Minkowski world. For the most part, we will equip the timelike hypersurface with a conformal metric and fix the trace of its extrinsic curvature, $K$. Since dS$_3$ and the three-dimensional flat cylinder are locally conformally equivalent, provided we retain the same value of $K$ we should view the two configurations as different states in the same theory. What distinguishes them is the remaining Cauchy (and corner) data along $\Sigma$, and, in turn, that they yield different values for the variant of the Brown--York boundary stress-energy tensor \cite{Odak:2021axr} pertinent to these boundary conditions. Our main finding is that there are no exponentially growing modes of the type uncovered in \cite{Anninos:2023epi,Liu:2024ymn}. The underlying reason for this is rooted in the structure of the constraint equation stemming from the projection of the time-time component of Einstein's field equations onto $\Gamma$. This reads
\begin{equation}
    \mathcal{D}_m\mathcal{D}^m \bomega + \frac{1}{2}\mathcal{D}_m \bomega \mathcal{D}^m \bomega - \frac{R^{(\Gamma)}}{4} - \frac{64 \pi^2G_N^2}{4}T^{mn}T_{mn}\, e^{-4\bomega} + \frac{K^2}{6}\,e^{2\bomega}=0 \, ,
    \label{eq: Weyl eom in}
\end{equation}
where $T_{mn}$ is a traceless analogue of the Brown--York stress tensor, $\bomega(x^m)$ is the dynamical Weyl factor of the boundary metric, and $R^
{(\Gamma)}$ its Ricci scalar. As we shall see, a basic property of a background with an umbilic boundary is that it yields a vanishing $T_{mn}$. As such, the linearisation of (\ref{eq: Weyl eom in}) leads to a decoupling of the linearised Weyl mode $\delta\bomega$ from the bulk dynamics, which would ordinarily be mediated through the $T^{mn}T_{mn}$ term. But then, we may hope that the exponentially growing linear modes are further tamed, since these came about precisely through the interaction of these two terms. We will find that this is indeed the case at the linearised level. From this point of view, small deviations from the umbilic state with $T_{mn}=0$ control the onset of the exponentially growing modes uncovered in \cite{Anninos:2023epi,Liu:2024ymn}.

The paper is structured as follows. In section \ref{sec:framework}, we review the general framework for the initial boundary value problem at hand. In section \ref{sec:linear}, we set up the linearised problem, and study the perturbations of the decoupled boundary Weyl mode $\bomega(x^m)$. We identify linear and non-linear solutions. In section \ref{sec: bulk}, we set up the linearised gravity problem for the bulk modes and establish the absence of exponentially growing modes whose imaginary frequency grows with $l$. In section \ref{sec: Dirichlet}, we analyse the Dirichlet problem with a fixed dS$_3$ boundary, and also find linear stability in the inner region. Technical details are provided in the various appendices. 

\section{General framework} \label{sec:framework}

We consider general relativity with vanishing cosmological constant in $(d+1)$ spacetime dimensions, with Lorentzian action
\begin{equation}
   I = \frac{1}{16 \pi G_N} \int_{\mathcal{M}} d^{d+1} x \sqrt{-\text{det} \, g_{\mu\nu}} \, R + \frac{1}{8 \pi G_N  d} \int_\Gamma d^{d}x \sqrt{-\text{det}\,g_{mn}} \, K  \, .
    \label{eq: action}
\end{equation}
Here $\Gamma$ is a timelike boundary with coordinates $x^m$ and induced metric $g_{mn}$, and $K = g^{mn} K_{mn}$ is the trace of the extrinsic curvature $K_{mn} = \tfrac{1}{2} \mathcal{L}_{n^\mu} g_{mn}$, with $n^\mu$ the outward-pointing unit normal. The boundary term, which differs from the Gibbons--Hawking--York term by a factor of $\tfrac{1}{d}$, results in a good variational problem  for conformal boundary conditions \cite{An:2021fcq,Odak:2021axr}.\footnote{The same boundary term is required to have a well-defined variational problem for the umbilic boundary conditions \cite{Fournodavlos:2021eye}. In this case, one fixes the umbilicity condition $K_{mn} = \alpha g_{mn}$, with $\alpha = \tfrac{K}{3}$ constant, instead of the conformal boundary conditions.} These fix the conformal class of the induced metric, $[g_{mn}]|_{\Gamma}$, and $K$ at $\Gamma$. Solutions obey the vacuum Einstein equations, 
\begin{equation} 
R_{\mu\nu} - \tfrac{1}{2}g_{\mu\nu} R = 0 \,.
\end{equation}
Writing the boundary metric as
\begin{equation}
    ds^2|_\Gamma = g_{mn} dx^m dx^n\equiv e^{2\bomega}[g]_{mn} dx^m dx^n \,,
\label{eq:bdrymet}\end{equation}
and varying the on-shell action with respect to the conformal representative $[g]_{mn}$ defines the conformal Brown--York stress tensor density \cite{An:2021fcq,Odak:2021axr},
\begin{equation}
    T_{m n} \equiv \left. -\frac{e^{(d-2)\bomega}}{8\pi G_N}\left( K_{m n} - \frac{1}{d} K g_{mn} \right) \right|_\Gamma  \,.
    \label{eq: tmn}
\end{equation}
It is traceless with respect to $[g]_{mn}$, and transforms with conformal weight $(2-d)$ under Weyl transformations of the conformal representative.

The projections of the Einstein equations onto $\Gamma$ take a simple form in terms of $T_{mn}$. The momentum constraint gives
\begin{equation}
\mathcal{D}^m T_{mn} =-\frac{(d-1)}{8\pi d G_{N}}\,e^{d\bomega} \mathcal{D}_n K\;,
\label{eq:momconst}\end{equation}
where $\mathcal{D}_m$ is the covariant derivative of $[g]_{mn}$, which is also used to raise and lower boundary indices. Hence, $T_{mn}$ is conserved when $K$ is constant. The Hamiltonian constraint along $\Gamma$ yields a second-order, non-linear equation for the Weyl factor,
\begin{equation}
    \mathcal{D}_m\mathcal{D}^m \bomega + \frac{d-2}{2}\mathcal{D}_m \bomega \mathcal{D}^m \bomega - \frac{R^{(\Gamma)}}{2(d-1)} - \frac{64 \pi^2G_N^2}{2(d-1)}T^{mn}T_{mn}\, e^{-2(d-1)\bomega} + \frac{K^2}{2d}\,e^{2\bomega}=0 \, ,
    \label{eq: Weyl eom}
\end{equation}
where $R^{(\Gamma)}$ is the Ricci scalar of $[g]_{mn}$. %

In this paper we focus on $d=3$. The background is empty four-dimensional Minkowski space, with a timelike boundary whose induced metric is that of three-dimensional de Sitter space, satisfying conformal boundary conditions.

\subsection{Background solution} \label{sec:background}

We start by considering four-dimensional Minkowski space in spherical coordinates,
\begin{equation}
    ds^2 = - dt^2 + dr^2 + r^2 d\Omega_2^2 \,, \qquad d\Omega_2^2 \equiv d\theta^2 + \sin^2 \theta\, d\varphi^2 \,,
    \label{eq: Minkowski}
\end{equation}
with $t\in\mathbb{R}$, $r\geq0$, $\theta\in[0,\pi]$ and $\varphi\in[0,2\pi)$. In the wedge $r>|t|$, outside the light cone of the event $O$ at $t=r=0$, we introduce coordinates $\tau\in\mathbb{R}$ and $\rho>0$ through
\begin{equation}
    t = \rho \sinh \tau \,, \qquad r = \rho \cosh \tau \,,
    \label{eq: dS coordinates}
\end{equation}
in terms of which the metric \eqref{eq: Minkowski} becomes
\begin{equation}
    ds^2 = d\rho^2 + \rho^2 \left( - d\tau^2 + \cosh^2 \tau \, d\Omega_2^2 \right) \,.
    \label{eq: dS foliation}
\end{equation}
The term in brackets is the metric of dS$_3$ with unit radius, such that surfaces of constant $\rho$ are the timelike hyperboloids $r^2-t^2=\rho^2$, with the induced metric of dS$_3$ with radius $\rho$. The wedge is bounded by the null rays $t=\pm r$, reached as $\rho\to0$ with $\tau\to\pm\infty$. The vector field $\partial_\tau$ generates the boosts preserving the wedge, whose bifurcation point is $O$, see figure~\ref{fig:background}.

We place $\Gamma$ at $\rho=\r$, dividing the wedge into an inner region, $\rho\in(0,\r]$, and an outer region, $\rho\in[\r,\infty)$. At $t=0$, $\Gamma$ is a round sphere of radius $\r$, and its radius $\r\cosh\tau=\sqrt{\r^2+t^2}$ grows exponentially in $\tau$. Each point of $\Gamma$ at fixed angles has constant proper acceleration $\tfrac{1}{\r}$. With boundary coordinates $x^m=(\tau,\theta,\varphi)$ and outward-pointing unit normal $n^\mu\partial_\mu=\sigma\,\partial_\rho$, where $\sigma=+1$ ($\sigma=-1$) for the inner (outer) region, we find
\begin{equation}
    \bar{g}_{mn} dx^m dx^n = \r^2\left(-d\tau^2+\cosh^2 \tau \, d\Omega_2^2\right) \,, \qquad \bar{K}_{mn} = \frac{\sigma}{\r}\, \bar{g}_{mn} \,, \qquad K = \frac{3\sigma}{\r} \, .
    \label{eq: induced}
\end{equation}
Barred quantities refer to the background metric. Thus, $K\in(0,\infty)$ for the inner region and $K\in(-\infty,0)$ for the outer region. As $\r\to\infty$, $|K|\to0$ and $\Gamma$ approaches spatial infinity $i^0$, such that the inner region fills the wedge while the outer region shrinks towards $i^0$. As $\r\to0$, $|K|\to\infty$ and $\Gamma$ collapses onto the null rays, such that the inner region shrinks towards $O$ while the outer region fills the wedge. What we mean here, when taking limits of dimensionful quantities, is as compared to some external reference scale like the energy of some experiment or the size of a galaxy.

Crucially, $\bar{K}_{mn}$ is pure trace, that is, $\Gamma$ is an umbilic surface. It follows from \eqref{eq: tmn} that
\begin{equation}
    \bar{T}_{mn} = 0 \,,
    \label{eq: Tmn zero}
\end{equation}
and consequently all the boundary charges built from $\bar{T}_{mn}$, including the conformal energy associated to the conformal Killing vector $\cosh\tau\,\partial_\tau$, vanish. Since $\bar{T}_{mn}=0$ and $R^{(\Gamma)}=\tfrac{6}{\r^2}$, the background with $\bar{\bomega}=0$ satisfies \eqref{eq: Weyl eom}.

The vanishing stress tensor contrasts with the analogous case with a spherical cavity $r=\r$ in Minkowski space \cite{Anninos:2023epi}, whose induced metric is
\begin{equation}
    ds^2|_\Gamma = -dt^2 + \r^2 d\Omega_2^2 \,,
    \label{eq: cavity metric}
\end{equation}
with $\bar{K}_{mn}dx^mdx^n = \r\, d\Omega_2^2$ and $K= \tfrac{2}{\r}$, such that $8\pi G_N \bar{T}_{tt} = -\tfrac{2}{3\r}\neq0$. Nevertheless, the induced metric on the $\mathrm{dS}_3$ boundary is conformal to a finite strip of the static cylinder. Indeed, introducing $\frakt$ through $\tan\frakt=\sinh\tau$, the induced metric \eqref{eq: induced} becomes
\begin{equation}
    ds^2\big|_\Gamma
    =
    \frac{\r^2}{\cos^2\frakt}
    \left(-d\frakt^2+d\Omega_2^2\right).
\end{equation}
This is Weyl equivalent to the static-cylinder metric \eqref{eq: cavity metric}, restricted to
$   \frakt=\frac{t}{\r}\in
    \left(-\frac{\pi}{2},\frac{\pi}{2}\right)$.

\begin{figure}[t!]
\centering
\begin{tikzpicture}[scale=1]
	\pgfmathsetmacro\L{3}      %
	\pgfmathsetmacro\rr{2}     %
	\pgfmathsetmacro\c{\L/180} %
	\coordinate (O)  at (0,0);
	\coordinate (ip) at (0,\L);
	\coordinate (im) at (0,-\L);
	\coordinate (i0) at (\L,0);
	\coordinate (A)  at (\L/2,\L/2);
	\coordinate (B)  at (\L/2,-\L/2);
	\fill[fill=blue, fill opacity=.4] (O) -- (B)
		-- plot[variable=\s, domain=-6:6, samples=120] ({\c*(atan(\rr*exp(\s))+atan(\rr*exp(-\s)))},{\c*(atan(\rr*exp(\s))-atan(\rr*exp(-\s)))})
		-- (A) -- cycle;
	\fill[fill=magenta, fill opacity=.4] (B)
		-- plot[variable=\s, domain=-6:6, samples=120] ({\c*(atan(\rr*exp(\s))+atan(\rr*exp(-\s)))},{\c*(atan(\rr*exp(\s))-atan(\rr*exp(-\s)))})
		-- (A) -- (i0) -- cycle;
	\foreach \tu in {-2,-1.5,-1,-0.5,0,0.5,1,1.5,2}{
		\draw[black!55, line width=0.3pt] (O)
			-- plot[variable=\u, domain=0.5:89.5, samples=80] ({\c*(atan(tan(\u)*exp(\tu))+atan(tan(\u)*exp(-\tu)))},{\c*(atan(tan(\u)*exp(\tu))-atan(tan(\u)*exp(-\tu)))})
			-- (i0);
	}
	\pgfmathsetmacro\xz{\c*2*atan(\rr)}
	\coordinate (P0) at (\xz,0);
	\draw[dashed] (ip) .. controls (0.3,\L-0.7) and (\xz,0.7) .. (P0);
	\draw[dashed] (im) .. controls (0.3,-\L+0.7) and (\xz,-0.7) .. (P0);
	\draw (B)
		-- plot[variable=\s, domain=-6:6, samples=120] ({\c*(atan(\rr*exp(\s))+atan(\rr*exp(-\s)))},{\c*(atan(\rr*exp(\s))-atan(\rr*exp(-\s)))})
		-- (A);
	\draw (im) -- (ip) -- (i0) -- cycle;
	\draw (O) -- (A);
	\draw (O) -- (B);
	\path (ip) -- (i0) node[pos=.25, above, sloped] {$\mathcal{I}^+$};
	\path (im) -- (i0) node[pos=.25, below, sloped] {$\mathcal{I}^-$};
	\path (O) -- (A) node[pos=.45, above, sloped] {\small $t=r$};
	\path (O) -- (B) node[pos=.45, below, sloped] {\small $t=-r$};
	\draw[->, shorten >=1pt] (\L*0.95,-\L*0.45) node[right] {\small $\rho=\r$} -- (1.99,-0.59);
	\draw[->, shorten >=1pt] (-0.35,\L*0.55) node[left] {\small $r=\r$} -- (0.59,2.18);
	\node[left] at (im) {$i^-$};
	\node[left] at (ip) {$i^+$};
	\node[right] at (i0) {$i^0$};
	\node[left] at (O) {$O$};
	\filldraw (O) circle (0.03cm);
	\filldraw (A) circle (0.03cm);
	\filldraw (B) circle (0.03cm);
\end{tikzpicture}
\caption{Radial Penrose diagram of Minkowski space, where each point represents a two-sphere and the vertical line on the left is $r=0$. The wedge $r>|t|$ between the null rays emanating from the event $O$ is foliated by the timelike hyperboloids $\rho=$ constant, each carrying a dS$_3$ induced metric. The umbilic boundary $\Gamma$ at $\rho=\r$ (solid curve) divides the wedge into the inner region (blue) and the outer region (magenta). All such boundaries end on $\mathcal{I}^\pm$ at the points where the null rays from $O$ reach null infinity. As $\r\to\infty$ the boundary approaches spatial infinity $i^0$, while as $\r\to0$ it collapses onto the null rays. The thin grey lines are slices of constant $\tau$, which run from $O$ to $i^0$. For comparison, the dashed curve shows, schematically, the static spherical boundary $r=\r$, which runs from $i^-$ to $i^+$ and meets $\Gamma$ on the $t=0$ slice.}
\label{fig:background}
\end{figure}

\section{Linearised  boundary dynamics} \label{sec:linear}

We now consider linearised perturbations first around an arbitrary background $\bar g_{\mu\nu}$,
\begin{equation}
    g_{\mu\nu} = \bar{g}_{\mu\nu} + \varepsilon \, h_{\mu\nu} \,, \;\;|\varepsilon| \ll1 \,.
    \label{eq: metric pert}
\end{equation}
We denote linearised variations by $\varepsilon \, \delta X \equiv X(\bar{g}+\varepsilon h) - X(\bar{g})$.  Choosing $[g]_{mn} = \bar{g}_{mn}$, such that $\bar{\bomega}=0$, the conformal boundary conditions read
\begin{equation}
     ds^2|_\Gamma = e^{2\bomega(x^{m})}\, \bar{g}_{mn}dx^m dx^n \,, \quad\quad \delta K|_\Gamma = 0 \,,
\label{eq:linCBCs}\end{equation}
where $\bomega = \varepsilon\, \delta\bomega$ is dynamical. We must also specify boundary conditions for the diffeomorphism parameter. At the linearised level, a vector field $\xi^\mu$ acts on the metric perturbation as
\begin{equation}
     h_{\mu\nu}
    \rightarrow
    h_{\mu\nu}
    +\bar\nabla_\mu\xi_\nu
    +\bar\nabla_\nu\xi_\mu.
\end{equation}

We refer to diffeomorphisms that preserve the conformal boundary conditions and satisfy
\begin{equation}
    \xi_\perp\big|_\Gamma=0,
    \qquad
    \xi_\perp\equiv \bar n_\mu\xi^\mu,
\end{equation}
as allowed diffeomorphisms, where $\bar n^\mu$ is the outward-pointing unit normal to $\Gamma$ in the background geometry. In the coordinates of \eqref{eq: dS foliation},
$
    \bar n^\mu\partial_\mu=\sigma\,\partial_\rho
$. They leave the location of $\Gamma$ fixed and are regarded as gauge redundancies. By contrast, a perturbation may be locally expressible as a diffeomorphism and satisfy the conformal boundary conditions while having $\xi_\perp|_\Gamma\neq0$. Such a perturbation  will be referred to as a physical diffeomorphism. In particular, physical diffeomorphisms may change the dynamical Weyl factor at the boundary.

The momentum constraint then implies $\mathcal{D}^m \delta T_{mn}=0$, while linearising \eqref{eq: Weyl eom} gives
\beq
\mathcal{D}_{m} \mathcal{D}^{m}\delta\bomega+\left( 64\pi^{2}G_{N}^{2}\,\bar{T}_{mn}\bar{T}^{mn}
+\frac{K^{2}}{d} \right)\delta\bomega = \frac{64\pi^{2}G_{N}^{2}}{d-1}\,\bar{T}^{mn}\delta T_{mn}\;,
\label{eq: linearised Weyl}
\eeq
with $\mathcal{D}_{m}$ the covariant derivative of $\bar{g}_{mn}$. The source term encodes the bulk dynamics, and $\delta T_{mn}$ itself contains $\delta\bomega$ through \eqref{eq: tmn}. 

For the background \eqref{eq: induced}, $\bar{T}_{mn}=0$, so the source term vanishes, even though $\delta T_{mn}$ itself need not. In both regions, the Weyl factor then obeys
\begin{equation}
    \mathcal{D}_m\mathcal{D}^m\delta\bomega + \frac{K^2}{3}\,\delta\bomega = 0 \,,
    \label{eq: Weyl dS3}
\end{equation}
with $\mathcal{D}_m$ the covariant derivative of the dS$_3$ metric \eqref{eq: induced}. That is, $\delta\bomega$ is effectively a free scalar field on dS$_3$ with mass squared $\mu_{\text{eff}}^2=-\tfrac{K^2}{3}=-\tfrac{3}{\r^2}$. Since $\delta T_{mn}$ does not appear, this equation can in principle be solved without any information from the bulk equations to linear order.

It is also important to note that \eqref{eq: Weyl dS3} governs physical diffeomorphisms normal to $\Gamma$. Since the background boundary is umbilic, a diffeomorphism that is purely in the orthogonal direction gives
\begin{equation}
    \delta_\xi g_{mn}
    =2\xi_\perp\bar K_{mn}
    =\frac{2K}{3}\,\xi_\perp\,\bar g_{mn},
    \qquad
    \delta_\xi\bomega
    =\frac{K}{3}\,\xi_\perp.
    \label{eq: normal diffeo Weyl}
\end{equation}
The remaining condition $\delta_\xi K|_\Gamma=0$ gives
\begin{equation}\label{eq: normal diffeo Weyl ii}
\left(\mathcal{D}_m\mathcal{D}^m+\tfrac{3}{\r^2}\right)\xi_\perp=0~, 
\end{equation}
which coincides with \eqref{eq: Weyl dS3}. Thus, every solution of \eqref{eq: Weyl dS3} has a locally diffeomorphic realisation with $\xi_\perp=\tfrac{3}{K}\,\delta\bomega$. A non-trivial $\xi_\perp$ cannot be removed by a pure diffeomorphism and therefore describes a physical perturbation. We note further that (\ref{eq: normal diffeo Weyl ii}) is the linearised dynamics of the conformal factor in three-dimensional general relativity with $\Lambda = +\tfrac{1}{\r^2}$. 

Equations such as (\ref{eq: Weyl dS3}) and (\ref{eq: normal diffeo Weyl ii}) naturally appear in the consideration of spatial infinity \cite{AshtekarHansen1978,Beig:1982ifu}, $i^0$, where $\xi_\perp$ subject to (\ref{eq: normal diffeo Weyl ii}) plays the role of an $i^0$ supertranslation generator, while the $sl(2,\mathbb{C})$ isometries of dS$_3$ are the Lorentz symmetries. It would be interesting to interpret, from this point of view, (\ref{eq: Weyl eom}) as a non-linear and finite extension of the equation governing the structure of $i^0$. The structure of spatial infinity \cite{AshtekarHansen1978,Beig:1982ifu} also includes a transverse tensor akin to $T_{mn}$, which can be taken to be traceless as a further condition \cite{Compere:2011db}. For a configuration built entirely from $\bomega$, one finds
\begin{equation}
     \delta T^{(\bomega)}_{mn}= \frac{1}{8\pi G_N}\frac{3}{K}\left(\mathcal{D}_m\mathcal{D}_n-\frac{1}{3}\,\bar g_{mn}\,\mathcal{D}^2\right)\delta\bomega \,.
\end{equation}
From the above equation it follows, as we will soon verify, that for those $\delta \bomega$ that result from the action of bulk Poincar\'e translations, $\delta T^{(\bomega)}_{mn}$ vanishes. For the remaining $\delta\bomega$ satisfying (\ref{eq: Weyl dS3}), this is not so. Instead, in the latter case one has two functions of $S^2$ worth of data corresponding to the two independent solutions of (\ref{eq: Weyl dS3}). 

\subsection{Linearised boundary Weyl mode} \label{sec: weyl}

We focus on the inner region, $\sigma=+1$, and comment on the outer region at the end of this subsection. Since \eqref{eq: Weyl dS3} is linear and the background \eqref{eq: dS foliation} is spherically symmetric, we can study each angular mode separately. We write $\delta\bomega=\delta\bomega_{lm}(\tau)\,\mathbb{S}$, where $\mathbb{S}=Y_{lm}(\theta,\varphi)$ are the scalar spherical harmonics, labelled by the total angular momentum $l=0,1,2,\ldots$ and the axial angular momentum $m=-l,\ldots,l$, and satisfying $\tilde{\nabla}^2\mathbb{S}=-l(l+1)\,\mathbb{S}$, with $\tilde{\nabla}^2$ the Laplacian on the unit two-sphere. Equation \eqref{eq: Weyl dS3} then becomes
\begin{equation}
    \partial_\tau^2\delta\bomega_{lm} + 2\tanh\tau\,\partial_\tau\delta\bomega_{lm} + \left(\frac{l(l+1)}{\cosh^2\tau}-3\right)\delta\bomega_{lm} = 0 \,.
    \label{eq: Weyl modes}
\end{equation}
The equation is identical for every $m$, so we suppress this label from now on. Moreover, $\bar{T}_{mn}=0$ implies that the bulk does not source the Weyl mode, whose evolution is therefore determined entirely by the initial data $\{\delta\bomega_l,\partial_\tau\delta\bomega_l\}$ at the corner $\tau=0$. For $l\geq2$, the solutions can be expressed in terms of associated Legendre functions of order two. This representation degenerates for $l<2$, so we treat the $l=0,1$ modes separately.

\textbf{Modes with $l=0,1$.} The general solutions are
\begin{equation}
    \delta\bomega_0 = a_0\sinh\tau + b_0\,\frac{\cosh 2\tau}{\cosh\tau} \,, \qquad \delta\bomega_1 = a_1\cosh\tau + b_1\,\frac{\sinh\tau\left(2\cosh^2\tau+1\right)}{\cosh^2\tau} \,,
    \label{eq: Weyl l01}
\end{equation}
where $a_l$ and $b_l$ are constants fixed by the initial data. All these solutions grow as $e^{|\tau|}$ at early and late times. %
The solutions multiplying $\{a_0,a_1\}$ have a simple interpretation in terms of translations in four-dimensional Minkowski space. In Cartesian coordinates, the boundary can be described by
\[
    x^\mu\big|_\Gamma = (t,x^i)|_\Gamma
    =
    \r\left(\sinh\tau,\cosh\tau\,\hat n^i\right), \qquad i=1,2,3\,,
\]
where $\hat n^i$ is a unit vector on $S^2$. For the inner region, the unit normal has components
$\bar n^\mu = \tfrac{x^\mu}{\r}$. A constant translation $\xi^\mu=c^\mu$ therefore has normal component
\begin{equation}
    \xi_\perp
    =\bar n_\mu c^\mu
    =-c^0\sinh\tau
     +c_i\hat n^i\cosh\tau.
    \label{eq: translation normal component}
\end{equation}
Using \eqref{eq: normal diffeo Weyl}, a temporal translation generates the $a_0$ solution, whereas the three spatial translations generate the $a_1$ solutions, since the components $\hat n^i$ are $l=1$ spherical harmonics. In appendix~\ref{app: l01}, we construct these diffeomorphisms explicitly, together with the Poincar\'e charges and the boundary charges carried by the $b_l$ modes. 

In appendix~\ref{sec: stress tensor}, we compute the linearised conformal stress tensor, and show that $\delta T^{(\bomega)}_{mn}=0$ under the four Poincar\'e translations. On the other hand, $\delta T^{(\bomega)}_{mn}$ is non-vanishing for the $l\ge 2$ modes. 

\textbf{Modes with $l\geq2$.} The general solution is
\begin{equation}
    \delta\bomega_l = \frac{1}{\cosh\tau}\Big(\alpha_l\,P_l^2(\tanh\tau)+\beta_l\,Q_l^2(\tanh\tau)\Big) \,,
    \label{eq: Weyl l2}
\end{equation}
where $P_l^2(z)$ and $Q_l^2(z)$ are the associated Legendre functions of the first and second kind, of degree $l$ and order two, and $\alpha_l$ and $\beta_l$ are constants fixed by the initial data. A related analysis of gravitational perturbations in the dS$_3$ slicing of Minkowski space appears in \cite{Liu:2021dfc}. The first branch decays as $e^{-3|\tau|}$, e.g.~$\tfrac{P_2^2(\tanh\tau)}{\cosh\tau}=\tfrac{3}{\cosh^3\tau}$, while the second grows as $e^{|\tau|}$. Note that the two branches have definite parity under the antipodal map of dS$_3$, $(\tau,\theta,\varphi)\to(-\tau,\pi-\theta,\varphi+\pi)$. The decaying $\alpha_l$ modes are even, while the growing $\beta_l$ modes are odd. Likewise, for $l=0,1$ the $b_l$ solutions in \eqref{eq: Weyl l01} are even and the $a_l$ solutions are odd. We come back to this when discussing the outer patch below.

All solutions of \eqref{eq: Weyl modes} therefore grow at most as $e^{|\tau|}$, for any $l$. Since $t=\r\sinh\tau$ along $\Gamma$, $e^{|\tau|}\simeq \tfrac{2|t|}{\r}$ at late times, such that the Weyl factor grows at most linearly in Minkowski time, at a rate independent of $l$. In particular, the Weyl sector admits no modes whose growth rate increases with $l$, in contrast with the spherical cavity, where modes growing as $e^{c\,l^{1/3}t}$ were found \cite{Anninos:2023epi,Liu:2024ymn,Liu:2025xij}. In the spherical case, these growing modes are sourced by the $\bar T^{mn} \delta T_{mn}$ term \cite{Anninos:2025zgr}, which vanishes here.

\textbf{The outer patch, briefly.} For the outer region, $\sigma=-1$, \eqref{eq: Weyl dS3} and its solutions are unchanged, since only $K^2$ enters. As $\r\to\infty$, however, the outer region shrinks onto spatial infinity, where it is natural to compare with the Beig--Schmidt description of asymptotically flat spacetimes \cite{Beig:1982ifu}. There, in the hyperbolic slicing \eqref{eq: dS foliation}, the leading correction to the metric, $g_{\rho\rho}=1+\tfrac{2\varsigma}{\rho}+\ldots$, is governed by a function $\varsigma(\tau,\theta,\varphi)$ on the unit dS$_3$ obeying the same equation, $(\mathcal{D}_m\mathcal{D}^m+3)\varsigma=0$. For Schwarzschild with mass $M$, $\varsigma=G_NM \tfrac{\cosh2\tau}{\cosh\tau}$, which is the $b_0$ solution in \eqref{eq: Weyl l01}. Asymptotic conditions that further impose the parity condition $\varsigma(\tau,\theta,\varphi)=\varsigma(-\tau,\pi-\theta,\varphi+\pi)$ \cite{Beig:1982ifu,Compere:2011db} remove the $a_{0}$, $a_1$, and $\beta_l$ solutions, including the translations. For $\delta\bomega$, this amounts to restricting the initial data to $\delta\bomega|_{\tau=0}$ even and $\partial_\tau\delta\bomega|_{\tau=0}$ odd under the antipodal map $(\theta,\varphi)\to(\pi-\theta,\varphi+\pi)$. For such data, the Weyl factor with $l\geq2$ decays as $e^{-3|\tau|}$.

\subsection{A non-linear remark}

It is also natural to ask about the non-linear equation that one finds upon setting ${T}_{mn}=0$ in (\ref{eq: Weyl eom}). Further setting $d=3$, this equation reads
\begin{equation}
   - \mathcal{D}_m\mathcal{D}^m \bomega - \frac{1}{2}\mathcal{D}_m \bomega \mathcal{D}^m \bomega + \frac{R^{(\Gamma)}}{4} = \frac{K^2}{6}\,e^{2\bomega} \, .
    \label{eq: Weyl nl eom}
\end{equation}
What the above equation states is that the Ricci scalar of a metric of the form $e^{2\bomega}\bar{g}_{mn}$ is constant and positive, specifically $\tfrac{2}{3}K^2$. This equation has many solutions, as we are not constraining anything else about the three-geometry. It is a Lorentzian version of the Yamabe equation \cite{ginoux2015lorentzian} that seeks a metric of constant Ricci scalar within a conformal class. 

Upon setting the total angular momentum to zero, we can find some explicit solutions. A non-linear completion of the $a_0$ mode in (\ref{eq: Weyl l01}) is given by
\begin{equation}
\bomega(\tau) = \log\frac{3}{K\frakr}  - \log\left(\cos\alpha-\sin\alpha\,\sinh\tau\right)~, \quad\quad \alpha \in \left(-\frac{\pi}{2},\frac{\pi}{2}\right)~.
\end{equation}
For small $\alpha$, we retrieve the perturbative solution. For $\alpha>0$, however, the argument of the logarithm vanishes at the finite time $\sinh\tau=\cot\alpha$. This corresponds to $\mathcal{I}^+$ of an asymptotically dS$_3$ solution in terms of the boundary three-metric. 

It is also possible to extend the $a_1$ modes in (\ref{eq: Weyl l01}) non-linearly. In this case, the Weyl factor becomes
\begin{equation}
   \bomega =\log\frac{3}{K\frakr}-\log\Big(\sqrt{1-p_0^2+p_ip^i}+p_0\sinh\tau+p_i\hat n^i\cosh\tau\Big) \,, \qquad |p_0|<1 \,,
\end{equation}
which reduces to the previous solution for $p_0=-\sin\alpha$ and $p_i=0$. At late times, the behaviour is analogous to the case with vanishing angular momentum. The Weyl factor behaves as  $\sim-\tau$ wherever $p_0+p_i\hat n^i>0$, while elsewhere the argument of the logarithm vanishes at finite $\tau$.

Returning to vanishing angular momentum, the non-linear completion of the $b_0$ mode can be written down in terms of elliptic-type integrals. What can be gleaned from this is that the largest late-time growth, at the non-linear level with vanishing angular momentum, is given by $\sim -\tau$. This is in contrast to the $\sim e^\tau$ growth at the linearised level. A limiting case of this family is
\begin{equation}
\bomega(\tau) = \log \frac{\sqrt{3}}{K\frakr} - \log \cosh {\tau}~,
    \label{eq: log cosh}
\end{equation}
which effectively maps the dS$_3$ boundary metric to a finite time interval of an Einstein static Universe. The $b_1$ modes are boosts of the $b_0$ mode, so their non-linear completion requires boosting these solutions, which also excites higher harmonics.

These solutions solve the boundary equation \eqref{eq: Weyl nl eom} alone, and require a vacuum infilling in which $\Gamma$ is umbilic. For the $a_0$ and $a_1$ completions, $e^{2\bomega}\bar g_{mn}$ is again dS$_3$ with radius $\tfrac{3}{K}$, and the infilling is Minkowski space. For the $b_0$ completion, it is Schwarzschild, by Birkhoff's theorem. The elliptic solutions are the distant umbilic surfaces (see footnote~\ref{fn: umbilic Sch}), while \eqref{eq: log cosh} is the part $|\frakt|<\tfrac{\pi}{2}$ of the photon sphere $r=3G_NM$, with $G_NM=\tfrac{1}{\sqrt{3}K}$. At linear order, $b_0=-\tfrac{G_NM}{\frakr}$, such that the energy $Q[\partial_t]$, defined in \eqref{eq: boundary charges}, vanishes.

These examples suggest a richer non-linear structure than is visible in the linearised theory. Understanding the evolution for general angular-dependent initial data, including the existence of globally regular solutions and the possible formation of finite-time conformal boundaries, remains an interesting open question. %

\section{Linearised bulk dynamics} \label{sec: bulk}

In this section we  set up the linearised perturbations in the bulk using the gauge-invariant method of Kodama and Ishibashi \cite{Kodama:2000fa,Kodama:2003jz}. For the background \eqref{eq: dS foliation}, this can be done in two ways. One may decompose the perturbations with respect to the dS$_3$ slices of constant $\rho$, in a $1+3$ split, or with respect to the round two-spheres, in a $2+2$ split. In the former, the harmonics live on the non-compact Lorentzian dS$_3$, and the boundary time $\tau$ does not appear explicitly. We therefore adopt the $2+2$ split, and write the background metric as
\begin{equation}
    ds^2 = g_{ab}\,dy^a dy^b + \rho^2\cosh^2\tau\, d\Omega_2^2 \,, \qquad g_{ab}\,dy^a dy^b = -\rho^2 d\tau^2 + d\rho^2 \,,
    \label{eq: 2+2}
\end{equation}
where $y^a=(\tau,\rho)$ are coordinates on the two-dimensional orbit space. The metric perturbation then decomposes into a scalar and a vector sector, $h_{\mu\nu}=h^{(S)}_{\mu\nu}+h^{(V)}_{\mu\nu}$. The scalar sector is built from the scalar harmonics $\mathbb{S}$ introduced in section~\ref{sec: weyl}, together with $\mathbb{S}_i=-\tfrac{1}{\sqrt{l(l+1)}}\tilde{\nabla}_i\mathbb{S}$, where $\tilde{\nabla}_i$ is the covariant derivative on the unit two-sphere. The vector sector is built from the transverse vector harmonics $\mathbb{V}_i=(\star d\mathbb{S})_i$, which satisfy $\tilde{\nabla}^i\mathbb{V}_i=0$ and $\left(\tilde{\nabla}^2+l(l+1)-1\right)\mathbb{V}_i=0$, for $l=1,2,\ldots$. Our conventions follow appendix C of \cite{Anninos:2023epi}.

Two caveats must be taken into account. First, the Kodama--Ishibashi master fields are only defined for $l\geq2$. Modes with $l=0,1$ carry no local gravitational degrees of freedom and must be treated separately. Second, the method is gauge invariant in the absence of boundaries. In the presence of $\Gamma$, only allowed diffeomorphisms are gauge redundancies, and the physical diffeomorphisms of section~\ref{sec: weyl}, with $\xi_\perp|_\Gamma\neq0$, are not captured by the master fields. The total perturbation is therefore
\begin{equation}
    h_{\mu\nu} = h^{(S)}_{\mu\nu} + h^{(V)}_{\mu\nu} + h^{(\text{diff})}_{\mu\nu} \,, \qquad h^{(\text{diff})}_{\mu\nu} = \bar{\nabla}_\mu\xi_\nu+\bar{\nabla}_\nu\xi_\mu \,, \qquad \xi_\perp|_\Gamma\neq0 \,.
    \label{eq: total pert}
\end{equation}
This is a key difference with the spherical cavity in Minkowski \cite{Anninos:2023epi}, dS$_4$ \cite{Anninos:2024wpy} and AdS$_4$ \cite{Anninos:2024xhc}, where the conformal boundary conditions force $\xi_\perp|_\Gamma=0$ for $l\geq2$, such that physical diffeomorphisms only appear for $l=0,1$. Here, since $\Gamma$ is umbilic, a normal displacement only changes the Weyl factor, and physical diffeomorphisms exist for every $l$, as shown in section~\ref{sec: weyl}. This makes the problem qualitatively different.

In what follows, we first discuss the modes with $l=0,1$. We then solve the bulk equations for $l\geq2$ using the Kodama--Ishibashi master fields, and finally impose the linearised conformal boundary conditions at $\Gamma$.

\subsection{Locally diffeomorphic modes with \texorpdfstring{$l\in \{0,1\}$}{l ∈ \{0,1\}}} \label{sec: l01}

For $l\in\{0,1\}$, every linearised vacuum perturbation can be locally expressed as a diffeomorphism. In the absence of a boundary, a distinguished subset of diffeomorphisms is given by the ten generators of the bulk Poincar\'e algebra. We now explore what remains of this in the presence of $\Gamma$. The Lorentz subalgebra, $sl(2,\mathbb{C})\cong so(1,3)$, is naturally associated to the isometries of the dS$_3$ boundary metric. More generally, dS$_3$ admits ten conformal Killing vectors. Six of these are the aforementioned Killing vectors, which are physical diffeomorphisms that do not vanish at $\Gamma$ and leave $\delta\bomega$ invariant. As such, the interior filling is invariant under the $sl(2,\mathbb{C})$.  The four spacetime translations of the Poincar\'e algebra result in a variation $\delta\bomega_l$, with $l\in\{0,1\}$, given by the $a_l$ terms discussed near \eqref{eq: Weyl l01}. These transform the boundary metric by the four non-isometric conformal Killing vectors of dS$_3$. 

Thus,  all ten Poincar\'e generators persist as physical diffeomorphisms in the presence of the finite boundary $\Gamma$ subject to conformal boundary conditions. 
For each of these, the corresponding Poincar\'e charge is given by a surface integral over a round sphere in the slice $\tau=0$. By the linearised constraints, this is independent of its radius. Shrinking the sphere to the vertex $O$, all charges vanish for initial data regular at $O$. In the inner region, the physical content of the modes with $l \in \{0,1\}$ is  given by the four $b_l$ modes. These carry the quasi-local energy and momentum of $\Gamma$ defined by the conformal Brown--York tensor \eqref{eq: tmn}. In the outer region, the charges are allowed, and require a compensating diffeomorphism to satisfy the conformal boundary conditions. %
We give more details and explicit expressions in appendix~\ref{app: l01}.

\subsection{Bulk linearised modes with \texorpdfstring{$l\geq 2$}{l ≥ 2}}

For $l\geq2$, the linearised Einstein equations are solved by $h^{(S)}_{\mu\nu}+h^{(V)}_{\mu\nu}$, with
\begin{equation}
\begin{cases}
    h^{(S)}_{\tau\tau} = \frac{1}{2}\left(\partial_\tau^2+\rho^2\partial_\rho^2-\rho\,\partial_\rho\right)\Phi^{(S)}\,\mathbb{S} \,, \\
    h^{(S)}_{\tau\rho} = \left(\partial_\tau\partial_\rho-\frac{1}{\rho}\partial_\tau\right)\Phi^{(S)}\,\mathbb{S} \,, \\
    h^{(S)}_{\rho\rho} = \frac{1}{2}\left(\partial_\rho^2-\frac{1}{\rho}\partial_\rho+\frac{1}{\rho^2}\partial_\tau^2\right)\Phi^{(S)}\,\mathbb{S} \,, \\
    h^{(S)}_{ij} = \frac{\rho^2\cosh^2\tau}{2}\left(\partial_\rho^2+\frac{1}{\rho}\partial_\rho-\frac{1}{\rho^2}\partial_\tau^2\right)\Phi^{(S)}\,\tilde{g}_{ij}\,\mathbb{S} \,,
\end{cases}
\qquad
\begin{cases}
    h^{(V)}_{\tau i} = -\rho\,\partial_\rho\Phi^{(V)}\,\mathbb{V}_i \,, \\
    h^{(V)}_{\rho i} = -\frac{1}{\rho}\partial_\tau\Phi^{(V)}\,\mathbb{V}_i \,,
\end{cases}
\label{eq: KI ansatz}
\end{equation}
and all other components vanishing. Here $\tilde{g}_{ij}$ is the metric on the unit two-sphere. 
In the notation of Kodama and Ishibashi, \eqref{eq: KI ansatz} corresponds to the gauge in which the scalar perturbation has no mixed components $h_{ai}$ and no traceless angular part, while the vector perturbation has no angular components $h_{ij}$. In this gauge, the non-vanishing components of $h_{\mu\nu}$ coincide with the gauge-invariant combinations of \cite{Kodama:2000fa,Kodama:2003jz}, which are determined by the master fields. The details are collected in appendix~\ref{app: KI}. Unlike the gauge used for the spherical cavity in \cite{Anninos:2023epi, Anninos:2024wpy, Anninos:2024xhc}, \eqref{eq: KI ansatz} is not adapted to $\Gamma$. A gauge in which $h^{(S)}_{\mu\nu}$ preserves the conformal class of the induced metric at $\Gamma$ would use the normal component of the gauge vector at $\Gamma$, and thereby absorb part of $h^{(\text{diff})}_{\mu\nu}$ into $h^{(S)}_{\mu\nu}$. With \eqref{eq: KI ansatz}, the normal diffeomorphisms are entirely contained in $h^{(\text{diff})}_{\mu\nu}$.

The master fields $\Phi^{(S/V)}(\tau,\rho)$ both satisfy
\begin{equation}
    \Box_2\left(\frac{\Phi^{(S/V)}}{\rho\cosh\tau}\right) - \frac{l(l+1)}{\rho^2\cosh^2\tau}\,\frac{\Phi^{(S/V)}}{\rho\cosh\tau} = 0 \,,
    \label{eq: master equation}
\end{equation}
where $\Box_2$ is the Laplacian of the orbit-space metric $g_{ab}$ in \eqref{eq: 2+2}, that is, of two-dimensional Minkowski space in Rindler coordinates. Explicitly,
\begin{equation}
    \rho^2\partial_\rho^2\Phi^{(S/V)} - \rho\,\partial_\rho\Phi^{(S/V)} - \partial_\tau^2\Phi^{(S/V)} + 2\tanh\tau\,\partial_\tau\Phi^{(S/V)} - \frac{l(l+1)-2}{\cosh^2\tau}\,\Phi^{(S/V)} = 0 \,.
    \label{eq: master equation explicit}
\end{equation}
Equivalently, since $\rho\cosh\tau=r$ is the Minkowski radius, $\Phi^{(S/V)}\,\mathbb{S}/r^2$ satisfies the massless wave equation in four-dimensional Minkowski space. Introducing $x=\log(\rho/\r)$, such that $\Gamma$ sits at $x=0$, \eqref{eq: master equation} becomes
\begin{equation}
    \partial_\tau^2\psi - \partial_x^2\psi + \frac{l(l+1)}{\cosh^2\tau}\,\psi = 0 \,, \qquad \psi \equiv \frac{\Phi^{(S/V)}}{\rho\cosh\tau} \,,
    \label{eq: PT}
\end{equation}
that is, a two-dimensional wave equation with a P\"oschl--Teller potential in time. In contrast with the spherical cavity, where the boundary is static and one can Fourier transform in time, here the potential depends on $\tau$, reflecting the time dependence of the dS$_3$ boundary. Instead, \eqref{eq: PT} is invariant under shifts of $x$ (dilatations of Minkowski space), such that separable solutions are of the form $e^{ikx}$ times a P\"oschl--Teller mode in $\tau$.

Equation \eqref{eq: PT} admits the energy
\begin{equation}
    E_{\text{bulk}}(\tau) = \frac{1}{2}\int dx\left[(\partial_\tau\psi)^2+(\partial_x\psi)^2+V\psi^2\right] \,, \qquad V\equiv\frac{l(l+1)}{\cosh^2\tau} \,,
    \label{eq: bulk energy}
\end{equation}
on the slices of constant $\tau$, which is non-negative. Up to boundary terms, it satisfies
\begin{equation}
    \frac{dE_{\text{bulk}}}{d\tau} = \frac{1}{2}\,\partial_\tau V\int dx\,\psi^2 = -\tanh\tau\int dx\,V\psi^2 \,,
    \label{eq: bulk energy derivative}
\end{equation}
and is therefore non-increasing for $\tau\geq0$. We say that a perturbation has finite energy if \eqref{eq: bulk energy} is finite. This will be a useful quantity to characterise the different perturbations.

Continuing, we can write the general linearised solution as a superposition of modes,
\begin{equation}
    \psi(\tau,x) = \int_{\mathcal{C}} dk\, e^{ikx}\Big(A(k)\,u_k(\tau)+B(k)\,u_{-k}(\tau)\Big) \,,
    \label{eq: Fourier}
\end{equation}
where $u_k(\tau)$ solves the P\"oschl--Teller equation
\begin{equation}
    \partial_\tau^2u_k + \left(k^2+\frac{l(l+1)}{\cosh^2\tau}\right)u_k = 0 \,,
    \label{eq: PT modes}
\end{equation}
and the contour $\mathcal{C}$ is discussed below. The coefficients $A(k)$ and $B(k)$ are fixed by the initial data $\{\psi,\partial_\tau\psi\}$ on the slice $\tau=0$, by Fourier transforming in $x$. These are in turn determined by the initial data of the metric perturbation, by inverting \eqref{eq: KI ansatz}. %

For integer $l$, the solutions of \eqref{eq: PT modes} are elementary functions. Explicitly, $u_k$ is the product of $e^{ik\tau}$ and a polynomial of degree $l$ in $\tanh\tau$, which can be generated recursively as
\begin{equation}
    u_k = \left(l\tanh\tau-\partial_\tau\right)\left((l-1)\tanh\tau-\partial_\tau\right)\cdots\left(\tanh\tau-\partial_\tau\right)e^{ik\tau} \,.
    \label{eq: PT recursion}
\end{equation}
For instance, for $l=2$,
\begin{equation}
    u_k = \left(3\tanh^2\tau-3ik\tanh\tau-1-k^2\right)e^{ik\tau} \,.
\end{equation}
For generic $k$, $u_k$ and $u_{-k}$ are two independent solutions of \eqref{eq: PT modes}. The exceptions are $k=0$ and $k=\pm ij$, with $j=1,\ldots,l$, where $u_k$ and $u_{-k}$ become proportional to each other. At these values, two independent solutions are given instead by the associated Legendre functions $P_l^{j}(\tanh\tau)$ and $Q_l^{j}(\tanh\tau)$, which solve \eqref{eq: PT modes} with $k^2=-j^2$, where $j=0$ corresponds to $k=0$.

The contour $\mathcal{C}$ depends on the behaviour of the solution in $x$, and there are two cases:
\begin{itemize}
    \item \textbf{Real $k$.} The modes oscillate in $x=\log(\tfrac{\rho}{\r})$, and $u_{\pm k}$ remain bounded in $\tau$. These describe perturbations with finite energy on the slices of constant $\tau$, for which $\mathcal{C}$ is the real line.
    \item \textbf{Imaginary $k=-i\kappa$}, with $\kappa$ real. The modes are power laws in $\rho$, $e^{ikx}=(\tfrac{\rho}{\r})^{\kappa}$, and generically grow as $e^{|\kappa||\tau|}$ at early and late times. The exceptions are $\kappa=\pm j$, with $j=1,\ldots,l$, for which a combination of $u_k$ and $u_{-k}$ decays as $\cosh^{-j}\tau$ at both ends. These are the bound states of the P\"oschl--Teller potential. Such contributions appear as discrete terms in \eqref{eq: Fourier}, or equivalently from deforming $\mathcal{C}$ away from the real axis.\footnote{{Differentiating with respect to $k$ gives further solutions, in which $e^{ikx}$ is multiplied by powers of $x=\log(\tfrac{\rho}{\r})$. We analyse these in section~\ref{sec: spectral}.}}
\end{itemize}

Near $\rho\to0$, where the slices of constant $\tau$ approach the event $O$, the modes with real $k$ oscillate increasingly fast in $x$, while the modes with $\kappa>0$ decay and those with $\kappa<0$ blow up. Near $\rho\to\infty$, where the slices approach spatial infinity, the situation is reversed. Thus, the bulk equations alone admit solutions that are ill-behaved at either end.

The discussion so far concerns the bulk equations alone. In the presence of the boundary $\Gamma$, only one of the two ends belongs to the region under consideration. For the inner region, only the behaviour as $\rho\to0$ is relevant. In particular, the modes with $\kappa>0$ decay there, so finite energy on the inner region does not exclude them. Since $e^{|\tau|}\simeq \tfrac{2|t|}{\r}$ at late times, these modes grow as $|t|^{\kappa}$ in Minkowski time. Although this growth is polynomial, $\kappa$ is not bounded by the bulk equations alone, such that the growth rate is not controlled. Which modes contribute, and for which values of $\kappa$, is determined by the conformal boundary conditions at $\Gamma$, to which we turn next.

\subsection{Conformal boundary conditions} \label{sec: CBC}

We will now proceed to impose the linearised conformal boundary conditions \eqref{eq:linCBCs} on the total perturbation \eqref{eq: total pert}, for $l\geq2$. These require that the induced metric at $\Gamma$ remains in the conformal class of $\bar{g}_{mn}$, that is,
\begin{equation}
    h_{mn}\big|_\Gamma = 2\,\delta\bomega\,\bar{g}_{mn} \,,
    \label{eq: conformal class}
\end{equation}
for some $\delta\bomega$, and that the trace of the extrinsic curvature is unchanged, $\delta K|_\Gamma=0$.

\textbf{Conformal class condition.} The condition \eqref{eq: conformal class} imposes that the perturbation does not modify the conformal class of the boundary metric. It factorises into independent conditions on each sector. First, since $\Gamma$ is umbilic, $h^{(\text{diff})}_{\mu\nu}$ only contributes to the Weyl factor through \eqref{eq: normal diffeo Weyl}. As such,  $h^{(\text{diff})}_{\mu\nu}$ does not receive any additional conditions from the imposition of \eqref{eq: conformal class}. Second, the mixed components of the scalar and vector sectors are proportional to $\mathbb{S}_i$ and $\mathbb{V}_i$, respectively, which are orthogonal on the two-sphere, such that \eqref{eq: conformal class} must hold separately in each sector. 

For the scalar sector, the mixed components $h^{(S)}_{\tau i}$ vanish and the angular components are pure trace in \eqref{eq: KI ansatz}. It then remains to require that the $\tau\tau$ and angular components carry the same Weyl factor, which gives
\begin{equation}
    \partial_\rho^2\Phi^{(S)}\big|_{\rho=\r} = 0  \,, \qquad \text{or} \qquad \left(\partial_x^2\psi+\partial_x\psi\right)|_{x=0}=0 \,.
    \label{eq: scalar BC}
\end{equation}
Using the bulk equation \eqref{eq: PT}, it can be written as a dynamical boundary condition
\begin{equation}
    \left.\left(\partial_\tau^2\psi + \frac{l(l+1)}{\cosh^2\tau}\,\psi + \partial_x\psi\right)\right|_{x=0} = 0 \,,
    \label{eq: scalar BC dynamical}
\end{equation}
which is second order in boundary time. For the vector sector, the angular components $h^{(V)}_{ij}$ vanish in \eqref{eq: KI ansatz}, and \eqref{eq: conformal class} requires $h^{(V)}_{\tau i}|_\Gamma=0$, that is,
\begin{equation}
    \partial_\rho\Phi^{(V)}\big|_{\rho=\r} = 0 \,, \qquad \text{or} \qquad \left(\partial_x\psi+\psi\right)\big|_{x=0} = 0 \,,
    \label{eq: vector BC}
\end{equation}
a Robin condition on $\psi$. %
It remains to impose $\delta K|_\Gamma=0$, which we do next.

\textbf{Mean curvature condition.} The variation of the trace of the extrinsic curvature at $\Gamma$ is
\begin{equation}
    2\,\delta K = \left.\left(\partial_\rho h^m{}_m - 2\mathcal{D}_m h_\rho{}^m - K h_{\rho\rho}\right)\right|_{\rho=\r} \,,
    \label{eq: deltaK}
\end{equation}
where boundary indices are raised with $\bar{g}_{mn}$. For the vector sector, $h^{(V)m}{}_m=h^{(V)}_{\rho\rho}=0$ and $\mathcal{D}_m h^{(V)}_\rho{}^m\propto\tilde{\nabla}^i\mathbb{V}_i=0$, such that $\delta K[h^{(V)}]=0$ identically, as for the spherical cavity \cite{Anninos:2023epi,Anninos:2024wpy,Anninos:2024xhc}. The scalar sector and the physical diffeomorphism, instead, both contribute to $\delta K$ and to the Weyl factor at $\Gamma$. Once \eqref{eq: scalar BC} holds, 
\begin{equation}
    h^{(S)}_{mn}|_\Gamma=2\,\delta\bomega^{(S)}\,\bar{g}_{mn}\,\mathbb{S} \,, \qquad \text{with} \qquad  \delta\bomega^{(S)}\equiv\tfrac{1}{4}\Box_2\Phi^{(S)}|_{\rho=\r} \,,
\end{equation} 
while \eqref{eq: normal diffeo Weyl} gives $\delta\bomega^{(\text{diff})}= \tfrac{K}{3}\,\xi_\perp$, writing $\xi_\perp=\xi_\perp(\tau)\,\mathbb{S}$. The Weyl factor analysed in section~\ref{sec: weyl} is then identified with the sum
\begin{equation}
    \delta\bomega_l = \delta\bomega^{(S)} + \delta\bomega^{(\text{diff})} \,.
    \label{eq: total Weyl}
\end{equation}

Evaluating \eqref{eq: deltaK}, we find two contributions to $\delta K$, each of the same form
\begin{equation}
    \delta K[h^{(S)}] = \frac{K}{3}\,\mathcal{O}_l\,\delta\bomega^{(S)}\,\mathbb{S} \,, \qquad \delta K[h^{(\text{diff})}] = \frac{K}{3}\,\mathcal{O}_l\,\delta\bomega^{(\text{diff})}\,\mathbb{S} \,,
    \label{eq: deltaK scalar}
\end{equation}
where $\mathcal{O}_l\equiv\partial_\tau^2+2\tanh\tau\,\partial_\tau+l(l+1)\cosh^{-2}\tau-3$ is the operator appearing in \eqref{eq: Weyl modes}. The condition $\delta K|_\Gamma=0$ is therefore precisely the Weyl mode equation \eqref{eq: Weyl modes} for the total Weyl factor \eqref{eq: total Weyl}, namely $\mathcal{O}_l\,\delta\bomega_l=0$. 

We note that the condition $\delta K|_\Gamma=0$ imposes no condition on the Kodama--Ishibashi bulk master field $\psi$ alone. Instead, given a $\delta\bomega_l$ solving \eqref{eq: Weyl modes} with some specified initial data (as analysed in section~\ref{sec: weyl}) and $\delta\bomega^{(S)}$ stemming from the scalar bulk perturbation which is fixed by its own initial data, the condition $\delta K|_\Gamma=0$ determines the physical diffeomorphism through a simple rearrangement of \eqref{eq: total Weyl},
\begin{equation}
    \delta\bomega^{(\text{diff})} = \delta\bomega_l - \delta\bomega^{(S)} \,.
    \label{eq: xi perp solution}
\end{equation}

\textbf{General linearised solution.} To summarise: for $l\geq2$, the general solution of the linearised Einstein equations satisfying the conformal boundary conditions is therefore
\begin{equation}
    h_{\mu\nu} = h^{(S)}_{\mu\nu}\big[\Phi^{(S)}\big] + h^{(V)}_{\mu\nu}\big[\Phi^{(V)}\big] + \bar{\nabla}_\mu\xi_\nu + \bar{\nabla}_\nu\xi_\mu \,, \qquad \xi_\perp\big|_\Gamma = \frac{3}{K}\left(\delta\bomega_l-\delta\bomega^{(S)}\right)\mathbb{S} \,,
    \label{eq: general solution}
\end{equation}
where the master fields solve \eqref{eq: master equation} subject to \eqref{eq: scalar BC} and \eqref{eq: vector BC}, and $\delta\bomega_l$ solves \eqref{eq: Weyl modes}. The free data are the bulk initial data of the two master fields, describing the two polarisations of the graviton, together with the corner data $\{\delta\bomega_l,\partial_\tau\delta\bomega_l\}$ of the boundary Weyl mode.

\textbf{Comparison with the spherical cavity.} The structure of \eqref{eq: general solution} differs qualitatively from the spherical cavity \cite{Anninos:2023epi,Anninos:2024wpy,Anninos:2024xhc}, as summarised in table~\ref{tab: comparison}. The key difference is the role of the physical diffeomorphism. In the spherical cavity, it is absent for $l\geq2$, and $\delta K[h^{(S)}]=0$ imposes a fourth-order condition on the scalar master field, which gives rise to the modes growing as $e^{c\,l^{1/3}t}$ \cite{Anninos:2023epi,Liu:2024ymn,Liu:2025xij}. %

Notably, since the spherical cavity admits no physical diffeomorphisms for $l \geq 2$, its corner initial data are not truly independent of the Cauchy data. Only the spherically symmetric mode $l = 0$ carries genuinely new corner data.\footnote{In the spherical cavity, the $l = 1$ mode corresponds to a rigid translation, carrying no physical corner data.} The umbilic boundary, by contrast, requires corner data for every $l$.

\begin{table}[h!]
\centering
\small
\begin{tabular}{lcc}
\toprule
 & Spherical cavity & Umbilic tube \\
\midrule
Background stress tensor $\bar{T}_{mn}$ & non-vanishing & vanishing \\
Physical diffeomorphisms with $l\geq2$ & absent & present \\
Scalar boundary condition & $\delta K=0$, fourth order & conformal class, second order \\
Vector boundary condition & conformal class, Robin & conformal class, Robin \\
Role of $\delta K=0$ & constrains $\Phi^{(S)}$ & determines $\xi_\perp$ \\
Weyl factor & determined by $\Phi^{(S)}$ & independent corner data \\
\bottomrule
\end{tabular}
\caption{Linearised conformal boundary conditions for $l\geq2$ in the spherical cavity and in the umbilic tube.}
\label{tab: comparison}
\end{table}

It remains to determine which bulk modes are compatible with the boundary conditions \eqref{eq: scalar BC} and \eqref{eq: vector BC}. We refer to this as the spectral problem. Inserting the general bulk solution \eqref{eq: Fourier} into the boundary conditions at $x=0$ relates the amplitudes of the modes with $k$ and $-k$, that is, it fixes how waves in $x$ are reflected at $\Gamma$. It also determines whether discrete modes with imaginary $k=-i\kappa$ and $\kappa>0$, which decay as $\rho\to0$ and grow as $|t|^\kappa$, are allowed in the inner region, and for which values of $\kappa$. A well-posed problem requires that such growth is absent, or at least that $\kappa$ remains bounded uniformly in $l$. We address this in the next subsection.

\subsection{Spectral problem} \label{sec: spectral}

We now determine which bulk modes are compatible with the boundary conditions \eqref{eq: scalar BC} and \eqref{eq: vector BC} in the inner region, $x\leq0$. We treat the scalar and vector sectors together. Since \eqref{eq: PT modes} only depends on $k^2$, the modes $e^{ikx}$ and $e^{-ikx}$ multiply the same solutions in $\tau$, and we consider
\begin{equation}
    \psi_k(\tau,x) = \left(a\,e^{ikx}+b\,e^{-ikx}\right)\Big(A\,u_k(\tau)+B\,u_{-k}(\tau)\Big) \,,
    \label{eq: mode}
\end{equation}
with constant $a$, $b$, $A$ and $B$.

\textbf{Real $k$.} The boundary conditions only act on the $x$-dependence of \eqref{eq: mode}. For the scalar sector, $\left(\partial_x^2\psi_k+\partial_x\psi_k\right)|_{x=0}=0$ gives
\begin{equation}
    \text{Scalar:} \quad \left(-k^2+ik\right)a + \left(-k^2-ik\right)b = 0 \,,
\end{equation}
while for the vector sector, $\left(\partial_x\psi_k+\psi_k\right)|_{x=0}=0$ gives
\begin{equation}
    \text{Vector:} \quad \left(1+ik\right)a + \left(1-ik\right)b = 0 \,.
\end{equation}
For $k\neq0$, the ratio $b/a$ defines the reflection coefficients
\begin{equation}
    \mathcal{R}_S(k) = \frac{1+ik}{1-ik} \,, \qquad \mathcal{R}_V(k) = -\frac{1+ik}{1-ik} \,, \qquad |\mathcal{R}_{S/V}(k)| = 1 \,.
    \label{eq: reflection}
\end{equation}
Thus, $\Gamma$ acts as a pure reflection, which relates the incoming and outgoing waves in $x$ by a phase. The reflection coefficients are independent of both $\tau$ and $l$, since the boundary conditions have constant coefficients in $x$. Moreover, for real $k$, $u_{\pm k}$ remain bounded in $\tau$.

\textbf{Imaginary $k$.} For $k=-i\kappa$, with $\kappa>0$, the two branches in \eqref{eq: mode} are $e^{\kappa x}$ and $e^{-\kappa x}$. Only the first one has finite energy as $\rho\to0$, such that $b=0$. The boundary conditions then require
\begin{equation}
\begin{cases}
    \kappa(\kappa+1) = 0 \,, \qquad & \text{for scalar sector,} \\ 
   \kappa+1 = 0 \,,  \qquad & \text{for vector sector,}
\end{cases}
\end{equation}
neither of which has a solution with $\kappa>0$. Therefore, the conformal boundary conditions remove all the discrete modes, which would grow as $|t|^\kappa$, for every $l$. 

Finally, at $k=0$ the two branches in $x$ degenerate into $1$ and $x$, neither of which has finite energy as $\rho\to0$.

\textbf{General solution.} It follows that, in the inner region, the general solution with finite energy is
\begin{equation}
    \psi(\tau,x) = \int_0^\infty dk\left(e^{ikx}+\mathcal{R}_{S/V}(k)\,e^{-ikx}\right)\Big(A(k)\,u_k(\tau)+B(k)\,u_{-k}(\tau)\Big) + \text{c.c.} \,,
    \label{eq: general psi}
\end{equation}
with $A(k)$ and $B(k)$ fixed by the initial data. Since $|\mathcal{R}_{S/V}|=1$ and $u_{\pm k}$ are bounded, no mode grows in time, for any $l$.

{\textbf{Logarithmic modes.} Differentiating \eqref{eq: mode} with respect to $k$ gives further solutions of \eqref{eq: PT},
\begin{equation}
    \psi=e^{\kappa x}\sum_{n=0}^{N}x^n f_n(\tau) \,,
    \label{eq: log modes}
\end{equation}
that is, powers of $\rho$ dressed by powers of $\log\rho$, the linearised counterpart of the logarithmic terms of \cite{Compere:2011db}. At $x=0$, the boundary conditions act on $x^ne^{\kappa x}$ as the $n$-th derivative of $\kappa(\kappa+1)$ and $\kappa+1$, respectively. Combined with \eqref{eq: PT}, this rules out modes with $\kappa\neq0$ and $N\geq1$, since both polynomials have simple roots, while modes with $\kappa=0$ have infinite energy. Logarithmic tails can still be present in the initial data, and are then carried by the continuum \eqref{eq: general psi}.}

{\textbf{The outer patch, briefly.} In the outer region, $x\geq0$, the boundary conditions are unchanged, but finite energy selects the branch $e^{-\kappa x}$, such that they become $\kappa(\kappa-1)=0$ and $\kappa-1=0$. Both are solved by $\kappa=1$. Hence, for every $l\geq2$ and in both sectors, the outer region supports the modes
\begin{equation}
    \psi=\frac{\r}{\rho}\Big(\alpha_l\,P_l^1(\tanh\tau)+\beta_l\,Q_l^1(\tanh\tau)\Big) \,.
    \label{eq: outer bound state}
\end{equation}
They have finite energy and are physical, but diverge as $\rho\to0$, so they are absent from the inner region. The $Q_l^1$ branch grows as $\cosh\tau$, %
at a rate independent of $l$.}

\subsection{Energy estimate} \label{sec: energy}

The absence of growth, uniformly in $l$, can be established directly for any finite-energy initial data. In the inner region, $x\leq0$, the bulk energy \eqref{eq: bulk energy} receives a flux $\partial_\tau\psi\,\partial_x\psi$ through $\Gamma$. Using \eqref{eq: scalar BC dynamical} and \eqref{eq: vector BC} to eliminate $\partial_x\psi$, this flux is a total $\tau$-derivative, up to a term proportional to $\partial_\tau V$. The total energy
\begin{equation}
    E^{(S/V)}(\tau) = E_{\text{bulk}}(\tau) + E^{(S/V)}_\Gamma(\tau) \,,
    \label{eq: energy}
\end{equation}
with the boundary energies
\begin{equation}
    E_\Gamma^{(S)} = \frac{1}{2}\left[(\partial_\tau\psi)^2+V\psi^2\right]_{x=0} \,, \qquad E_\Gamma^{(V)} = \frac{1}{2}\,\psi^2\big|_{x=0} \,.
    \label{eq: boundary energy}
\end{equation}
It follows that the energy satisfies
\begin{equation}
    \frac{dE^{(S)}}{d\tau} = \frac{1}{2}\,\partial_\tau V\left(\int_{-\infty}^{0}dx\,\psi^2 + \psi^2\big|_{x=0}\right) \,, \qquad \frac{dE^{(V)}}{d\tau} = \frac{1}{2}\,\partial_\tau V\int_{-\infty}^{0}dx\,\psi^2 \,.
    \label{eq: energy derivative}
\end{equation}
Both $E^{(S/V)}$ are sums of squares, since $V\geq0$, and are therefore non-negative. Since $\partial_\tau V=-2\tanh\tau\,V\leq0$ for $\tau\geq0$, it follows that
\begin{equation}
    E^{(S/V)}(\tau)\leq \alpha_l\,E^{(S/V)}(0) \,, \qquad \alpha_l=1 \,, \qquad \tau\geq0 \,,
    \label{eq: energy bound}
\end{equation}
and similarly for $\tau\leq0$. A mode growing as $e^{\kappa\tau}$ would make $E^{(S/V)}$ grow as $e^{2\kappa\tau}$, and is therefore excluded. The constant $\alpha_l$ measures the maximal amplification of the energy by the evolution. That $\alpha_l=1$ for every $l$ means that no angular momentum is amplified more than any other. Note that even though the initial energy $E^{(S/V)}(0)$ itself depends on $l$ through $V(0)=l(l+1)$, it is fixed by the initial data.

The energy bounds the first derivatives of $\psi$. The metric perturbation follows from $\Phi^{(S/V)}=\rho\cosh\tau\,\psi$ through \eqref{eq: KI ansatz}, which involves at most two derivatives of $\Phi^{(S/V)}$, with coefficients independent of $l$. The angular momentum can only enter through $V$, when \eqref{eq: PT} is used to trade $\partial_\tau^2\psi$ for $\partial_x^2\psi-V\psi$, and therefore polynomially. Hence, there is no room for growth rates that increase with $l$. This is in contrast with the spherical cavity, where the fourth-order boundary condition leads to an indefinite boundary energy, and the modes growing as $e^{c\,l^{1/3}t}$ require $\alpha_l$ to grow without bound with $l$.

{This analysis does not include the outer region, where the flux through $\Gamma$ enters with the opposite sign. The total energy is then no longer positive definite, which allows the growing modes \eqref{eq: outer bound state}.}

\subsection{Late-time behaviour} \label{sec: late}

The energy estimate guarantees that no perturbation grows. It is nevertheless instructive to describe the late-time behaviour explicitly. We summarise it here, and derive it in appendix~\ref{app: late} for {initial data regular at $O$}.

\textbf{Master fields.} At late times, $V\to0$, and \eqref{eq: PT} becomes the free wave equation
\begin{equation}
    \partial_\tau^2\psi-\partial_x^2\psi=0 \,,
    \label{eq: late wave}
\end{equation}
with general solution $\psi=F(\tau+x)+G(\tau-x)$. The first term is a wave travelling towards $x\to-\infty$, which leaves the inner region through the null cone $t=r$, while the second travels towards $\Gamma$. A wave $G$ reaching $\Gamma$ at late times $\tau$ must have been located at $x\simeq-\tau$ at early times. %
{For initial data regular at $O$, $\Phi^{(S/V)}\mathbb{S}/r^2$ is a regular solution of the wave equation, such that $\psi=O(e^{(l+1)x})$ as $x\to-\infty$. Such a wave is then suppressed as $e^{-(l+1)\tau}$, while earlier waves are reflected into waves $F$.}\footnote{{Finite energy alone allows $\psi\sim e^{\kappa x}$ for any $\kappa>0$, that is, initial data singular at $O$. For $0<\kappa<1$, the Weyl factor at $\Gamma$ then grows as $t^{1-\kappa}$.}} Hence, at late times, $\psi\simeq F(\tau+x)$ near $\Gamma$. 

At late times, the boundary conditions \eqref{eq: scalar BC dynamical} and \eqref{eq: vector BC} become
\begin{equation}
    \begin{split}
    \text{Scalar:} \quad & 0=\left(\partial_\tau^2\psi+V\psi+\partial_x\psi\right)\big|_\Gamma\simeq\left(\partial_\tau^2\psi+\partial_\tau\psi\right)\big|_\Gamma \,, \\
    \text{Vector:} \quad & 0=\left(\partial_x\psi+\psi\right)\big|_\Gamma\simeq\left(\partial_\tau\psi+\psi\right)\big|_\Gamma \,,
    \end{split}
\end{equation}
which are now ordinary differential equations for $\psi|_\Gamma$. The scalar one admits a constant solution, which is a remnant of the perturbation at $\Gamma$, while the vector condition forces $\psi|_\Gamma$ to decay as $e^{-\tau}$. Including the corrections from $V\simeq4l(l+1)e^{-2\tau}$, as derived in appendix~\ref{app: late}, one finds
\begin{equation}
    \psi\big|_\Gamma \simeq
    \begin{cases}
        u_\infty + \left(a_S+b_S\,\tau\right)e^{-\tau} \,, & \text{scalar,} \\
        \left(a_V+b_V\,\tau\right)e^{-\tau} \,, & \text{vector,}
    \end{cases}
    \label{eq: late psi}
\end{equation}
where the constants are determined by the initial data, while the rates are independent of $l$. For special initial data, $u_\infty$ vanishes. This is the case, for instance, for time-symmetric initial data with even $l$. 

Given this behaviour of $\psi$, the master field $\Phi^{(S/V)}$ at $\Gamma$ grows at most linearly in Minkowski time in the scalar sector, and logarithmically in the vector sector.

\textbf{Weyl factor.} Using \eqref{eq: scalar BC dynamical}, the Weyl factor induced by the scalar sector at $\Gamma$ can be written as
\begin{equation}
    \delta\bomega^{(S)} = -\frac{\cosh\tau}{4\r}\left(2\,\partial_\tau^2\psi + 2\tanh\tau\,\partial_\tau\psi + V\psi\right)\Big|_\Gamma \,.
    \label{eq: Weyl S boundary}
\end{equation}
Inserting \eqref{eq: late psi}, the constant $u_\infty$ and the terms proportional to $a_S$ cancel, and $\delta\bomega^{(S)}\to \tfrac{b_S}{4\r}$. For vanishing Weyl data, \eqref{eq: xi perp solution} then implies that the normal displacement $\xi_\perp=-\tfrac{3}{K}\,\delta\bomega^{(S)}$ approaches a constant. Thus, the boundary is permanently displaced by the passing radiation, which is reminiscent of a memory effect.

The total Weyl factor \eqref{eq: total Weyl} also contains $\delta\bomega_l$, the solution of \eqref{eq: Weyl modes} fixed by the Weyl data $\{\delta\bomega_l,\partial_\tau\delta\bomega_l\}$ at the corner $\tau=0$, which is not sourced by the bulk. Its late-time behaviour follows from section~\ref{sec: weyl}, and is summarised in table~\ref{tab: late Weyl}, together with $\delta\bomega^{(S)}$. The only growing modes are the branch $Q_l^2$ and the physical $b_l$ modes. By \eqref{eq: normal diffeo Weyl}, they describe a normal displacement of $\Gamma$ growing linearly in $t$, at a rate independent of $l$. For conformal boundary conditions, this implies that the linearised approximation breaks down for $t\gtrsim \tfrac{\r}{\varepsilon}$. For Dirichlet boundary conditions, these modes are absent, see section~\ref{sec: Dirichlet}. %

We illustrate these results in appendix~\ref{app: pulse}, where we evolve compactly supported initial data numerically. %

\begin{table}[H]
\centering
\small
\begin{tabular}{lcc}
\toprule
Mode & $l$ & Late times \\
\midrule
$\delta\bomega^{(S)}$ & $l\geq2$ & $\frac{b_S}{4\r}$ \\
$a_l$ & $l=0,1$ &  $e^{\tau}\simeq \frac{2t}{\r}$ \\
$b_l$ & $l=0,1$ & $e^{\tau}\simeq \frac{2t}{\r}$ \\
$P_l^2$ & $l\geq2$ & $e^{-3\tau}\simeq\left(\frac{\r}{2t}\right)^3$ \\
$Q_l^2$ & $l\geq2$ & $e^{\tau}\simeq \frac{2t}{\r}$ \\
\bottomrule
\end{tabular}
\caption{Different mode contributions to the late-time behaviour of the Weyl factor at $\Gamma$ { in the inner region}.}
\label{tab: late Weyl}
\end{table}

\section{Dirichlet boundary conditions} \label{sec: Dirichlet}

In this section, we replace the conformal boundary conditions by Dirichlet boundary conditions, which fix the induced metric at $\Gamma$.\footnote{The corresponding variational problem uses the Gibbons--Hawking--York term, that is, the boundary term in \eqref{eq: action} without the factor $\tfrac{1}{d}$.} At the linearised level, we find that homogeneous Dirichlet conditions imply the conformal ones, such that the Dirichlet solutions are precisely the solutions of section~\ref{sec: CBC} with vanishing Weyl factor. The bulk dynamics is therefore unchanged, while the Weyl modes of section~\ref{sec: weyl} are absent and the normal displacement of $\Gamma$ is fixed by the bulk. Inhomogeneous Dirichlet data are not obstructed, but lose one derivative with respect to conformal data.

\textbf{Dirichlet conditions imply conformal ones.} Homogeneous Dirichlet boundary conditions read
\begin{equation}
    h_{mn}\big|_\Gamma = 0 \,,
    \label{eq: Dirichlet}
\end{equation}
that is, \eqref{eq: conformal class} with $\delta\bomega=0$. They do not fix $\delta K$. However, linearising the Hamiltonian constraint \eqref{eq: Weyl eom} without setting $\delta K=0$, and using $\bar{T}_{mn}=0$ and $\bar K=\tfrac{3\sigma}{\r}$, gives
\begin{equation}
    \mathcal{D}_m\mathcal{D}^m\delta\bomega + \frac{3}{\r^2}\,\delta\bomega + \frac{K}{3}\,\delta K = 0 \,.
    \label{eq: Hamiltonian Dirichlet}
\end{equation}
For conformal boundary conditions, $\delta K=0$ and \eqref{eq: Hamiltonian Dirichlet} reduces to \eqref{eq: Weyl dS3}. For Dirichlet boundary conditions, $\delta\bomega=0$ and \eqref{eq: Hamiltonian Dirichlet} implies $\delta K=0$. Every solution of \eqref{eq: Dirichlet} therefore satisfies the conformal boundary conditions \eqref{eq:linCBCs} with $\delta\bomega=0$. This relies on $\Gamma$ being umbilic, such that $\bar T_{mn}=0$, and on $\bar K\neq0$.

\textbf{Modes with $l\geq2$.} The master fields obey the same boundary conditions \eqref{eq: scalar BC} and \eqref{eq: vector BC}, since these follow from \eqref{eq: conformal class}. The spectral problem of section~\ref{sec: spectral}, the energy estimate of section~\ref{sec: energy} and the late-time behaviour \eqref{eq: late psi} of the master fields are therefore unchanged. The only difference is that the total Weyl factor \eqref{eq: total Weyl} vanishes, $\delta\bomega_l=0$. Using \eqref{eq: normal diffeo Weyl}, this fixes the physical diffeomorphism in terms of the scalar master field,
\begin{equation}
    \xi_\perp\big|_\Gamma = -\frac{3}{K}\,\delta\bomega^{(S)}\,\mathbb{S} \,.
    \label{eq: xi perp Dirichlet}
\end{equation}
The general solution is thus \eqref{eq: general solution} with $\delta\bomega_l=0$. Its free data are the bulk initial data of the two master fields, while the corner data $\{\delta\bomega_l,\partial_\tau\delta\bomega_l\}$ of the conformal problem are absent. In particular, the normal displacements growing as $e^{|\tau|}$ in table~\ref{tab: late Weyl} do not appear {in the inner region, while in the outer region the scalar modes \eqref{eq: outer bound state} give a displacement growing as $t^2$}. In the inner region, \eqref{eq: xi perp Dirichlet} approaches the constant $-b_S/4$ at late times, see section~\ref{sec: late}.

\textbf{Modes with $l=0,1$.} The diffeomorphisms \eqref{eq: l01 diffeo} satisfy \eqref{eq: Dirichlet} if and only if $\delta\bomega_l=0$, that is, $b_l=0$ and $a_l=-c_l$. These are the translations and the Lorentz transformations, which are isometries of the background, such that $h_{\mu\nu}=0$. The compensating diffeomorphisms \eqref{eq: l01 charges} were chosen such that $\delta\bomega_l=0$, and therefore also satisfy \eqref{eq: Dirichlet}. Thus, the boundary charges \eqref{eq: boundary charges} reduce to the Poincar\'e charges. In the inner region, they vanish, and the modes with $l=0,1$ reduce to these Poincar\'e transformations, which carry no charge.

\textbf{Inhomogeneous data.} General Dirichlet data also change the conformal representative, $[g]_{mn}=\bar g_{mn}+\delta[g]_{mn}$, and \eqref{eq: Hamiltonian Dirichlet} becomes
\begin{equation}
    \delta K = \frac{3}{K}\left(\frac{1}{4}\,\delta R^{(\Gamma)} - \mathcal{D}_m\mathcal{D}^m\delta\bomega - \frac{3}{\r^2}\,\delta\bomega\right) \,,
    \label{eq: deltaK Dirichlet}
\end{equation}
where $\delta R^{(\Gamma)}$ is the linearised Ricci scalar of $[g]_{mn}$. Hence, the Dirichlet data are not constrained, but determine $\delta K$. Since \eqref{eq: deltaK Dirichlet} contains two derivatives of the Dirichlet data, while $\delta K$ contains one derivative of the metric, Dirichlet data of a given regularity determine $\delta K$ with one derivative less than conformal data of the same regularity. In this sense, the Dirichlet problem loses one derivative with respect to the conformal one. This is consistent with \cite{An:2025gvr}, where the non-linear Dirichlet problem is shown to be locally well-posed, despite such a loss of derivatives, provided $\bar K\bar g_{mn}-\bar K_{mn}$ is a Lorentzian metric. Here, $\bar K\bar g_{mn}-\bar K_{mn}=\tfrac{2K}{3}\,\bar g_{mn}$, so it indeed has Lorentzian signature.  

\begin{center}
\pgfornament[height=5pt, color=black]{83}
\end{center}
\vspace{5pt}

We end with some more general speculative remarks. Perhaps, to make sense of the Lorentzian conformal boundary conditions \cite{An:2021fcq} one should seek to identify a bulk configuration that permits an umbilic timelike boundary and perturb around that. When and whether such configurations exist within a given conformal class of boundary metrics, and for a given $K$, is not known to us.\footnote{\label{fn: umbilic Sch}One can also check that the Schwarzschild metric admits an umbilic boundary near its spatial infinity, in addition to the photon sphere which is a static umbilic surface at $r=3G_NM$ \cite{Claudel:2000yi,Cederbaum:2019rbv}. For the ordinary Schwarzschild coordinate system, the distant umbilic boundary satisfies the equation $\dot{r}^2 = {f(r)}^2 \left(1-\frac{9}{K^2 r^2}{f(r)}\right)$, with $f(r)=1-\frac{2G_NM}{r}$, and $K$ is related to the minimum radius of the surface $r_0>3G_NM$ as $K=\tfrac{3}{r_0}\sqrt{1-\tfrac{2G_NM}{r_0}}$. The induced metric at the umbilic boundary is locally conformally equivalent to dS$_3$. On the other hand a Kerr black hole does not admit a far away timelike umbilic surface. This can potentially be amended by the addition of gravitational radiation.}  A striking example where all physical configurations admit an umbilic boundary is the class of asymptotically AdS$_4$ spacetimes.

It is worth re-emphasising that the dS$_3$ boundary studied in this article seems to tie to the boundary structure at spatial infinity of an asymptotically Minkowski spacetime \cite{AshtekarHansen1978,Sommers1978,Beig:1982ifu,deBoer:2003vf, Raclariu:2021zjz}. In particular, the Lorentz algebra is associated to the $sl(2,\mathbb{C})$ isometries of dS$_3$, the infinitesimal Poincar\'e translations yield a vanishing $\delta T^{(\bomega)}_{mn}$, and the remaining physical diffeomorphisms (\ref{eq: Weyl dS3}) constitute two functions on $S^2$  worth of Weyl maps on dS$_3$. This is reminiscent of the $\text{Spi}$ group $SL(2,\mathbb{C}) \,{\rtimes} \, C^\infty(\text{dS}_3)$, subject to the Beig--Schmidt asymptotic equation (\ref{eq: Weyl dS3}). In light of our analysis, then, the asymptotic structure at $i^0$ naturally lends itself to a more finite realisation. The large number of physical diffeomorphisms preserving the umbilic condition, namely $\xi_\perp$ satisfying (\ref{eq: normal diffeo Weyl ii}), indicates that there may be an infinite symmetry algebra (or some approximate such structure near umbilic configurations) for the finite boundary also.

Echoing ideas in \cite{Banihashemi:2024yye,Banihashemi:2025qqi,Allameh:2025gsa}, given the link between umbilic boundaries and the vanishing of $T_{mn}$, one might be tempted to also consider the possibility of imposing the condition $T_{mn}=0$ as a constraint  \cite{Fournodavlos:2021eye} on the stress-energy tensor of the putative boundary dual. It is also interesting that the dS$_3$ umbilic boundary we studied has the property that a light signal travelling along the boundary reaches neighbouring points faster than a light signal that penetrates the bulk. Perhaps for such a state, the putative boundary theory has improved causality features. We note, however, that umbilicity alone does not guarantee the absence of bulk shortcuts. Moreover, beyond the linearised order there is no reason the boundary will retain umbilicity. Establishing the general conditions for which this can occur, along the lines of \cite{Gao:2000ga}, seems like an interesting problem in general relativity.

From a Euclidean perspective, improved causality properties might resonate with hints of locality appearing in thermodynamic expressions \cite{Anninos:2023epi,Banihashemi:2025qqi,minwalla}. In the case at hand, the Euclidean boundary would be an $S^3$. At the saddle point level, the Euclidean partition function yields (see equation (1.5) of \cite{Anninos:2024xhc})
\begin{equation}
\log  \mathcal{Z}_{\text{grav}}[S^3,K] \approx \frac{9\pi}{4G_N K^2} = \frac{\pi \mathfrak{r}^2}{4G_N}~,
\end{equation}
along with quantum corrections. The above expression is precisely equal to the Bekenstein--Hawking entropy, $\tfrac{A}{4G_N}$, of the dS$_3$ horizon radially emanating from the boundary into the bulk. That the logarithm of the partition function yields an entropy alone, rather than a free energy, is compatible with the vanishing of $T_{mn}$ for the umbilic boundary.\footnote{By Alexandrov's theorem \cite{Alexandrov:1962}, a closed, smooth, and connected   hypersurface embedded in a flat $\mathbb{R}^4$ with constant $K$ must be the round three-sphere. Thus, for a flat $\mathbb{R}^4$ filling the conformal boundary conditions alone enforce umbilicity, and hence $T_{mn}=0$. Moreover, for $K=0$ where the boundary round three-sphere lies at infinity, the flat filling is the unique saddle. This follows from the fact that a complete, Ricci-flat, asymptotically Euclidean four-manifold is flat by the Bishop--Gromov volume comparison \cite{Bishop:1964,Gromov:1981}, or alternatively by the positive action theorem \cite{Schoen:1979}.} Perhaps we should view this quantity as akin to the holographic $F$ \cite{Jafferis:2011zi,Casini:2012ei,Casini:2011kv} for a $\Lambda=0$ quantum Universe \cite{raclariu}. 

\noindent\section*{Acknowledgements}

We are grateful to Kuroush Allameh, Michael Anderson, Abhay Ashtekar, Silvia Georgescu, Juan Maldacena, Chawakorn Maneerat, K\'evin Nguyen, Ana-Maria Raclariu, Edgar Shaghoulian, Eva Silverstein, and Toby Wiseman for useful discussions. D.A. is funded by the Royal Society under the grant ``Concrete Calculables in Quantum de Sitter''. The research of D.A. is also supported in part by KU Leuven grant C16/25/010. R.A. is supported by CONICET and UNLP, Argentina. The work of D.A.G. is funded in part by UKRI Stephen Hawking Fellowship EP/W005530/1 ``Quantum Emergence of an Expanding Universe''. D.A. and D.A.G. are further supported by STFC consolidated grant ST/X000753/1. 

\appendix 

\section{Modes with \texorpdfstring{$l=0,1$}{l = 0, 1}} \label{app: l01}

In this appendix, we provide further details for subsection~\ref{sec: l01}. Without loss of generality, we take $m=0$, such that scalar perturbations are proportional to $\cos\theta$ and vector perturbations to $\sin^2\theta\,d\varphi$.

\textbf{Diffeomorphisms.} For $l=0$, the most general diffeomorphism can be written as
\begin{equation}
    \xi=\xi^\tau\partial_\tau+\xi^\rho\partial_\rho \,.
\end{equation}
The induced metric is conformal to $\bar{g}_{mn}$ if and only if $\partial_\tau(\xi^\tau/\cosh\tau)=0$, while $\delta K=0$ gives $\mathcal{O}_0\,\xi^\rho=0$, where $\mathcal{O}_l$ denotes the operator on the left-hand side of \eqref{eq: Weyl modes}. 

For $l=1$, a scalar diffeomorphism 
\begin{equation}
\xi=\cos\theta\left(\xi^\tau\partial_\tau+\xi^\rho\partial_\rho\right)+\xi^\theta\sin\theta\,\partial_\theta \,,
\end{equation}
preserves the conformal class if 
\begin{equation}
    \begin{cases}
       0 = \xi^\tau+\cosh^2\tau\,\partial_\tau\xi^\theta \,, \\
        0 = \xi^\theta - \partial_\tau\xi^\tau + \tanh\tau\,\xi^\tau \,,
    \end{cases}
\end{equation}
while $\delta K=0$ gives $\mathcal{O}_1\,\xi^\rho=0$. Thus, at $\Gamma$,
\begin{align}
    l=0:\quad & \xi^\tau = c_0\cosh\tau \,, \qquad \xi^\rho = \r\left(a_0\sinh\tau+b_0\,\frac{\cosh2\tau}{\cosh\tau}\right) \,, \nonumber\\
    l=1:\quad & \xi^\tau=c_1\sinh\tau+e_1 \,, \qquad \xi^\theta=\frac{c_1}{\cosh\tau}-e_1\tanh\tau \,, \nonumber\\ & \xi^\rho=\r\left(a_1\cosh\tau+b_1\,\frac{\sinh\tau\left(2\cosh^2\tau+1\right)}{\cosh^2\tau}\right) \,,
    \label{eq: l01 diffeo}
\end{align}
with
\begin{equation}
    \delta\bomega_0 = \left(a_0+c_0\right)\sinh\tau + b_0\,\frac{\cosh2\tau}{\cosh\tau} \,, \qquad \delta\bomega_1=\left(a_1+c_1\right)\cosh\tau+b_1\,\frac{\sinh\tau\left(2\cosh^2\tau+1\right)}{\cosh^2\tau} \,,
\end{equation}
in both regions. The vectors $\cosh\tau\,\partial_\tau$ and $\sinh\tau\cos\theta\,\partial_\tau+\tfrac{\sin\theta\,\partial_\theta}{\cosh\tau}$ are conformal Killing vectors of dS$_3$, which shift $a_0$ and $a_1$. The time translation $\partial_t=\tfrac{\cosh\tau}{\rho}\partial_\tau-\sinh\tau\,\partial_\rho$ corresponds to $a_0=-c_0=-\tfrac{1}{\r}$, and the spatial translation $\partial_z$ to $a_1=-c_1=\tfrac{1}{\r}$, both with $\delta\bomega_l=0$ and $h_{\mu\nu}=0$. The constant $e_1$ corresponds to the boost $z\partial_t+t\partial_z$, which is tangent to $\Gamma$. In the vector sector, $\xi=\xi^\varphi\,\partial_\varphi$ is tangent to $\Gamma$, and the conformal boundary conditions require $\partial_\tau\xi^\varphi=0$, such that the only vector diffeomorphisms are rotations.

\textbf{Poincar\'e charges.} In isotropic Cartesian coordinates $x^\mu=(t,\vec{x})$, with $r=|\vec{x}|$, the linearised Poincar\'e charges correspond to
\begin{equation}
    h_{\mu\nu}dx^\mu dx^\nu = 2\Phi\left(dt^2+d\vec{x}^{\,2}\right) - \frac{8G_N}{r}\left(P_i+\frac{\epsilon_{ijk}J^jx^k}{2r^2}\right)dt\,dx^i \,,
    \label{eq: l01 pert}
\end{equation}
with $\Phi=\tfrac{G_NM}{r}+\tfrac{G_N\left(D_i+t\,P_i\right)x^i}{r^3}$, that is, linearised Schwarzschild, boosted and displaced, and linearised Kerr. These perturbations change the conformal class of $\Gamma$. The conformal boundary conditions are restored by the compensating diffeomorphisms
\begin{align}
    M:\quad & \xi^\tau = \frac{2G_NM}{\r}\,\tau\cosh\tau \,, \qquad \xi^\rho = -G_NM\left(2\tau\sinh\tau+\frac{1}{\cosh\tau}\right) \,, \nonumber\\
    D:\quad & \xi^\rho=\frac{G_ND}{\r}\,\frac{\cosh2\tau}{\cosh^2\tau} \,, \qquad \xi^\theta=-\frac{2G_ND}{\r^2} \,, \nonumber\\
    P:\quad & \xi^\rho=G_NP\left(2\tau\cosh\tau+\frac{\sinh\tau\cosh2\tau}{\cosh^2\tau}\right) \,, \nonumber\\ & \xi^\tau=-\frac{2G_NP}{\r}\left(\tau\sinh\tau+\cosh\tau\right) \,, \qquad \xi^\theta=-\frac{2G_NP\,\tau}{\r\cosh\tau} \,, \nonumber\\
    J:\quad & \xi^\varphi=\frac{2G_NJ}{\r^2}\tanh\tau \,,
    \label{eq: l01 charges}
\end{align}
which are unique up to \eqref{eq: l01 diffeo} and rotations, and which we have chosen such that $\delta\bomega_l=0$. Thus, the charges do not source the Weyl factor. For $M$ and $P$, the normal displacement grows as $\tau e^{\tau}$. Relative to the areal radius of $\Gamma$, it grows logarithmically in $t$, such that the linearised description holds for $G_NM\,\tau,\,G_NP\,\tau\ll\r$. For $D$, the normal displacement is bounded, and $J$ only requires a rotation of the boundary frame.

\textbf{Boundary charges.} When $\delta K=0$, the linearised conformal Brown--York tensor $\delta T_{mn}$ is traceless and, by the momentum constraint \eqref{eq:momconst}, conserved. For any conformal Killing vector $\zeta^m$ of dS$_3$, the charge
\begin{equation}
    Q[\zeta] = \int_{S^2} d^2\Omega\,\r^2\cosh^2\tau\;\delta T_{mn}\,\zeta^m u^n \,, \qquad u^n\partial_n=\frac{1}{\r}\,\partial_\tau \,,
    \label{eq: boundary charge}
\end{equation}
is therefore independent of $\tau$. Since the conformal Killing vectors have $l\leq1$, only the modes with $l=0,1$ contribute. With the normal $\partial_\rho$, and for the tangential parts of $\partial_t$, $\partial_z$, the boost $z\partial_t+t\partial_z$ and the rotation $\partial_\varphi$, we find
\begin{equation}
    Q[\partial_t] = M + \frac{\r\,b_0}{G_N} \,, \qquad Q[\partial_z] = -P_z + \frac{\r\,b_1}{G_N} \,, \qquad Q[z\partial_t+t\partial_z] = D_z \,, \qquad Q[\partial_\varphi] = -J_z \,,
    \label{eq: boundary charges}
\end{equation}
while the $a_l$ modes, which are induced by the translations up to a conformal Killing vector of dS$_3$, carry no charge. The boundary charges thus reproduce the Poincar\'e charges, up to conventional signs, and the physical diffeomorphisms $b_0$ and $b_1$ contribute to the energy and momentum. In the inner region, where the Poincar\'e charges vanish, the energy and momentum of $\Gamma$ are entirely carried by the $b_l$ modes. In the outer region, the outward normal is $-\partial_\rho$, which flips the overall sign of \eqref{eq: boundary charges}.

\section{Linearised conformal stress tensor} \label{sec: stress tensor}

We now compute the linearised conformal stress tensor \eqref{eq: tmn} for the general solution of section~\ref{sec: CBC}, with conformal representative $[g]_{mn}=\bar g_{mn}$. %
Since $\bar T_{mn}=0$, the variation of the prefactor $e^{\bomega}$ in \eqref{eq: tmn} does not contribute. Moreover, since $h_{mn}|_\Gamma=2\,\delta\bomega\,\bar g_{mn}$ and $\delta K=0$, the variation of $K g_{mn}$ is pure trace. Hence, $\delta T_{mn}$ is determined by the traceless part of $\delta K_{mn}$,
\begin{equation}
    8\pi G_N\,\delta T_{mn}=-\left(\delta K_{mn}-\frac{1}{3}\,\bar g_{mn}\,\bar g^{pq}\,\delta K_{pq}\right)\bigg|_\Gamma \,,
    \label{eq: deltaT TF}
\end{equation}
which is traceless with respect to $\bar g_{mn}$ and, by the momentum constraint \eqref{eq:momconst}, conserved.

\textbf{Weyl factor.} Each term in the general solution \eqref{eq: general solution} contributes separately to \eqref{eq: deltaT TF}. A normal displacement $\xi_\perp$ of $\Gamma$ in flat space changes the extrinsic curvature by
\begin{equation}
    \delta K_{mn}=-\mathcal{D}_m\mathcal{D}_n\xi_\perp+\xi_\perp\,\bar K_{mp}\bar K^p{}_n \,.
    \label{eq: deltaK normal}
\end{equation}
Since $\Gamma$ is umbilic, the second term is pure trace, such that the physical diffeomorphism only contributes through the traceless part of $\mathcal{D}_m\mathcal{D}_n\xi_\perp$. Writing $\xi_\perp=\xi_\perp(\tau)\,\mathbb{S}$, as in section~\ref{sec: CBC}, \eqref{eq: normal diffeo Weyl} and \eqref{eq: xi perp solution} give
\begin{equation}
    \xi_\perp=\frac{3}{K}\left(\delta\bomega_l-\delta\bomega^{(S)}\right) ,
    \label{eq: xi perp K}
\end{equation}
where $\delta\bomega_l$ is the total Weyl factor \eqref{eq: total Weyl}, and $\delta\bomega^{(S)}$ is the Weyl factor induced by the scalar master field. We group the part proportional to $\delta\bomega_l$ into a Weyl contribution, and the part proportional to $\delta\bomega^{(S)}$ with the scalar sector. In terms of $\delta\bomega=\delta\bomega_l\,\mathbb{S}$, this gives
\begin{equation}
    \delta T_{mn}=\delta T^{(\bomega)}_{mn}+\delta T^{(S)}_{mn}+\delta T^{(V)}_{mn} \,,
    \label{eq: deltaT decomposition}
\end{equation}
with
\begin{equation}
    8\pi G_N\,\delta T^{(\bomega)}_{mn}=\frac{3}{K}\left(\mathcal{D}_m\mathcal{D}_n-\frac{1}{3}\,\bar g_{mn}\,\mathcal{D}^2\right)\delta\bomega \,,
    \label{eq: deltaT Weyl covariant}
\end{equation}
while $\delta T^{(S/V)}_{mn}$ only depend on the master fields at $\Gamma$, and are given below. The Weyl term \eqref{eq: deltaT Weyl covariant} is conserved precisely because of the Weyl equation \eqref{eq: Weyl dS3}. In components, and using \eqref{eq: Weyl dS3},
\begin{equation}
\begin{cases}
    8\pi G_N\,\delta T^{(\bomega)}_{\tau\tau}=\dfrac{3}{K}\left(\partial_\tau^2-1\right)\delta\bomega=\dfrac{3}{K}\left(2-2\tanh\tau\,\partial_\tau+\dfrac{\tilde\nabla^2}{\cosh^2\tau}\right)\delta\bomega \,, \\[3mm]
    8\pi G_N\,\delta T^{(\bomega)}_{\tau i}=\dfrac{3}{K}\left(\partial_\tau-\tanh\tau\right)\tilde\nabla_i\,\delta\bomega \,, \\[3mm]
    8\pi G_N\,\delta T^{(\bomega)}_{ij}=\dfrac{3}{K}\left[\tilde\nabla_i\tilde\nabla_j-\dfrac{1}{2}\,\tilde g_{ij}\tilde\nabla^2+\dfrac{\cosh^2\tau}{2}\,\tilde g_{ij}\left(\partial_\tau^2-1\right)\right]\delta\bomega \,.
\end{cases}
\label{eq: deltaT Weyl}
\end{equation}
As for the spherical cavity in dS$_4$ \cite{Anninos:2025zgr}, the components are expressed directly in terms of the Weyl factor. Here, however, \eqref{eq: deltaT Weyl} is exact at the linearised level, for any $K$ and every $l$.

For the modes \eqref{eq: Weyl l2} with $l\geq2$, we write $\delta\bomega_l=\Pi^2_l/\cosh\tau$, with
\begin{equation}
    \Pi^m_l\equiv\alpha_l\,P^m_l(\tanh\tau)+\beta_l\,Q^m_l(\tanh\tau) \,, \qquad P^m_l(z)\equiv\left(1-z^2\right)^{m/2}\partial_z^mP_l(z) \,,
    \label{eq: Pi}
\end{equation}
and similarly for $Q^m_l$. Using the recurrence relations of the associated Legendre functions, \eqref{eq: deltaT Weyl} becomes
\begin{equation}
\begin{cases}
    8\pi G_N\,\delta T^{(\bomega)}_{\tau\tau}=\dfrac{3}{K}\,(l-1)l(l+1)(l+2)\,\dfrac{\Pi^0_l}{\cosh^3\tau}\,\mathbb{S} \,, \\[3mm]
    8\pi G_N\,\delta T^{(\bomega)}_{\tau i}=-\dfrac{3}{K}\,(l-1)(l+2)\,\dfrac{\Pi^1_l}{\cosh^2\tau}\,\tilde\nabla_i\mathbb{S} \,, \\[3mm]
    8\pi G_N\,\delta T^{(\bomega)}_{ij}=\dfrac{3}{K}\left[\dfrac{\Pi^2_l}{\cosh\tau}\left(\tilde\nabla_i\tilde\nabla_j-\dfrac{1}{2}\,\tilde g_{ij}\tilde\nabla^2\right)\mathbb{S}+\dfrac{(l-1)l(l+1)(l+2)}{2}\,\dfrac{\Pi^0_l}{\cosh\tau}\,\tilde g_{ij}\,\mathbb{S}\right] .
\end{cases}
\label{eq: deltaT Weyl modes}
\end{equation}

\textbf{Modes with $l=0,1$.} There are no master fields. In the inner region, where the Poincar\'e charges vanish, $\delta T_{mn}=\delta T^{(\bomega)}_{mn}$, while in the outer region the charges of appendix~\ref{app: l01} contribute in addition. The $a_l$ modes in \eqref{eq: Weyl l01} are restrictions to $\Gamma$ of linear functions of the Minkowski coordinates, for which $\mathcal{D}_m\mathcal{D}_n\delta\bomega$ is proportional to $\bar g_{mn}$. Thus, they do not contribute to $\delta T_{mn}$, despite growing as $e^{|\tau|}$. For the $b_0$ mode, with $\mathbb{S}=1$, \eqref{eq: deltaT Weyl} and \eqref{eq: boundary charge} give
\begin{equation}
    8\pi G_N\,\delta T_{\tau\tau}=\frac{6}{K}\,\frac{b_0}{\cosh^3\tau} \,, \qquad Q[\partial_t]=\frac{3\,b_0}{G_NK} \,,
    \label{eq: b0 stress}
\end{equation}
reproducing \eqref{eq: boundary charges}.

\textbf{Bulk contributions.} For $l\geq2$, let $\psi^{(S/V)}_\Gamma(\tau)\equiv\psi^{(S/V)}(\tau,x=0)$ be the master fields at $\Gamma$. Using \eqref{eq: scalar BC dynamical} and \eqref{eq: vector BC} to eliminate the normal derivatives, the vector sector gives
\begin{equation}
\begin{cases}
    8\pi G_N\,\delta T^{(V)}_{\tau\tau}=0 \,, \\[3mm]
    8\pi G_N\,\delta T^{(V)}_{\tau i}=\dfrac{\sigma}{2}\,(l-1)(l+2)\,\dfrac{\psi^{(V)}_\Gamma}{\cosh\tau}\,\mathbb{V}_i \,, \\[3mm]
    8\pi G_N\,\delta T^{(V)}_{ij}=-\dfrac{\sigma}{2}\,\partial_\tau\left(\cosh\tau\,\psi^{(V)}_\Gamma\right)\left(\tilde\nabla_i\mathbb{V}_j+\tilde\nabla_j\mathbb{V}_i\right) .
\end{cases}
\label{eq: deltaT vector}
\end{equation}
This has the same structure as for the spherical cavity \cite{Anninos:2025zgr}. Although vector perturbations do not change the Weyl factor, they carry momentum density and shear. For the scalar sector, we find
\begin{equation}
\begin{cases}
    8\pi G_N\,\delta T^{(S)}_{\tau\tau}=\dfrac{\sigma}{4}\,(l-1)l(l+1)(l+2)\,\dfrac{\psi^{(S)}_\Gamma}{\cosh^3\tau}\,\mathbb{S} \,, \\[3mm]
    8\pi G_N\,\delta T^{(S)}_{\tau i}=-\dfrac{\sigma}{4}\,(l-1)(l+2)\,\dfrac{\partial_\tau\psi^{(S)}_\Gamma}{\cosh\tau}\,\tilde\nabla_i\mathbb{S} \,, \\[3mm]
    8\pi G_N\,\delta T^{(S)}_{ij}=-\dfrac{3}{K}\,\delta\bomega^{(S)}\left(\tilde\nabla_i\tilde\nabla_j-\dfrac{1}{2}\,\tilde g_{ij}\tilde\nabla^2\right)\mathbb{S}+\dfrac{\sigma}{8}\,(l-1)l(l+1)(l+2)\,\dfrac{\psi^{(S)}_\Gamma}{\cosh\tau}\,\tilde g_{ij}\,\mathbb{S} \,,
\end{cases}
\label{eq: deltaT scalar}
\end{equation}
where, by \eqref{eq: Weyl S boundary},
\begin{equation}
    -\frac{3}{K}\,\delta\bomega^{(S)}=\frac{\sigma}{4}\left[2\,\partial_\tau\left(\cosh\tau\,\partial_\tau\psi^{(S)}_\Gamma\right)+\frac{l(l+1)}{\cosh\tau}\,\psi^{(S)}_\Gamma\right] .
    \label{eq: Weyl S K}
\end{equation}
Both \eqref{eq: deltaT vector} and \eqref{eq: deltaT scalar} are traceless and conserved for arbitrary $\psi^{(S/V)}_\Gamma$. The vector part is the stress tensor of $h^{(V)}_{\mu\nu}$ alone, which does not change the Weyl factor. The scalar part is the stress tensor of $h^{(S)}_{\mu\nu}$ dressed by the normal displacement $-\tfrac{3}{K}\,\delta\bomega^{(S)}$, which removes its Weyl factor. Both perturbations have $\delta\bomega=\delta K=0$, and therefore solve the Dirichlet problem of section~\ref{sec: Dirichlet}.

We end this appendix with a few remarks. First, for bulk initial data supported away from $\Gamma$, as in appendix~\ref{app: pulse}, $\delta T^{(S/V)}_{mn}$ vanish at $\tau=0$, such that $\delta T_{mn}=\delta T^{(\bomega)}_{mn}$ there. Then \eqref{eq: deltaT Weyl} gives
\begin{equation}
    8\pi G_N\,\delta T_{\tau\tau}\big|_{\tau=0}=\frac{3}{K}\left(2+\tilde\nabla^2\right)\delta\bomega\big|_{\tau=0} \,, \qquad 8\pi G_N\,\delta T_{\tau i}\big|_{\tau=0}=\frac{3}{K}\,\tilde\nabla_i\,\partial_\tau\delta\bomega\big|_{\tau=0} \,.
    \label{eq: corner stress}
\end{equation}
Thus, the corner data $\{\delta\bomega_l,\partial_\tau\delta\bomega_l\}$ are the initial energy and momentum densities of the boundary.

Second, for the spherical cavity, the components of the scalar stress tensor can be written entirely in terms of the Weyl factor \cite{Anninos:2025zgr}. This relies on $\bar T_{mn}\neq0$, such that $\bar T^{mn}\delta T_{mn}$ sources the Weyl equation, and on the absence of physical diffeomorphisms for $l\geq2$. Here, $\bar T_{mn}=0$ decouples the bulk from the Weyl factor, and the bulk contributions \eqref{eq: deltaT scalar} are independent data. The Dirichlet problem, with $\delta\bomega=0$ but $\delta T_{mn}\neq0$, illustrates this.

Finally, in an orthonormal frame of $\bar g_{mn}$, and in the inner region, all contributions to $\delta T_{mn}$ decay at late times. Using \eqref{eq: late psi}, the bulk contributions decay at least as $e^{-2\tau}$. The growing modes of the Weyl factor, $Q^2_l$ and $b_l$ in table~\ref{tab: late Weyl}, have a linearised stress tensor that decays at least as $e^{-\tau}$. In terms of Minkowski time along $\Gamma$,
\begin{equation}
    e^{-\tau}\simeq\frac{3}{2Kt} \,.
\end{equation}
Interestingly, the boundary stress tensor does not grow, even when the Weyl factor does.

\section{Kodama--Ishibashi formalism} \label{app: KI}

In this appendix, we review the Kodama--Ishibashi formalism \cite{Kodama:2000fa,Kodama:2003jz}, following appendix C of \cite{Anninos:2023epi}, and use it to obtain the linearised solution \eqref{eq: KI ansatz}. The background \eqref{eq: 2+2} is of the form
\begin{equation}
	ds^2 = g_{ab}\,dy^a dy^b + r^2\, \tilde{g}_{ij}\,dz^i dz^j \,, \qquad g_{ab}\,dy^a dy^b = -\rho^2 d\tau^2 + d\rho^2 \,, \qquad r = \rho\cosh\tau \,,
	\label{eq: KI background}
\end{equation}
where $y^a=(\tau,\rho)$, $z^i=(\theta,\varphi)$, and $\tilde{g}_{ij}$ is the metric on the unit two-sphere. The orbit space $g_{ab}$ is the Rindler wedge of two-dimensional Minkowski space, with Minkowski coordinates $(t,r)=(\rho\sinh\tau,\rho\cosh\tau)$. We denote by $D_a$ its covariant derivative, by $\Box_2=D^aD_a$ its Laplacian, and by $\epsilon_{ab}$ its volume form, with $\epsilon_{\rho\tau}=\rho$.

\textbf{Metric perturbation.} Any metric perturbation can be uniquely decomposed into vector ($V$) and scalar ($S$) perturbations, $h_{\mu\nu}=h^{(V)}_{\mu\nu}+h^{(S)}_{\mu\nu}$, where
\begin{equation}
\begin{cases}
    h_{ab}^{(V)} = 0 \,, \\
    h_{ai}^{(V)} = r f^V_a \,\mathbb{V}_i \,, \\
    h_{ij}^{(V)} = 2r^2 H^V \,\mathbb{V}_{ij} \,,
\end{cases}
\qquad \qquad
\begin{cases}
    h_{ab}^{(S)} = f_{ab}\,\mathbb{S} \,, \\
    h_{ai}^{(S)} = r f^S_a \,\mathbb{S}_i \,, \\
    h_{ij}^{(S)} = 2r^2 H^S \,\mathbb{S}_{ij} + 2r^2\gamma \,\tilde{g}_{ij}\,\mathbb{S} \,.
\end{cases}
\label{eq: KI decomposition}
\end{equation}
The coefficients $f_{ab}$, $f_a^{V/S}$, $H^{V/S}$ and $\gamma$ are functions of $y^a$. The harmonics $\mathbb{S}$, $\mathbb{S}_i$ and $\mathbb{V}_i$ are defined in section~\ref{sec: bulk}, with eigenvalues $k_S^2=l(l+1)$ and $k_V^2=l(l+1)-1$, and the symmetric traceless tensors are
\begin{equation}
    \mathbb{S}_{ij} \equiv \frac{1}{k_S^2}\,\tilde{\nabla}_i\tilde{\nabla}_j\mathbb{S} + \frac{1}{2}\,\tilde{g}_{ij}\,\mathbb{S} \,, \qquad \mathbb{V}_{ij} \equiv -\frac{1}{2k_V}\left(\tilde{\nabla}_i\mathbb{V}_j+\tilde{\nabla}_j\mathbb{V}_i\right) .
\end{equation}
For $l=1$, both vanish identically, and $\mathbb{V}_i$ is a Killing vector of the two-sphere. We therefore restrict to $l\geq2$.

\textbf{Gauge invariants.} A diffeomorphism can be decomposed in the same way, $\xi_\mu=\xi^{(V)}_\mu+\xi^{(S)}_\mu$, with
\begin{equation}
\begin{cases}
    \xi_a^{(V)} = 0 \,, \\
    \xi_i^{(V)} = r\mathcal{L}^V\,\mathbb{V}_i \,,
\end{cases}
\qquad \qquad
\begin{cases}
    \xi_a^{(S)} = \mathcal{T}_a\,\mathbb{S} \,, \\
    \xi_i^{(S)} = r\mathcal{L}^S\,\mathbb{S}_i \,,
\end{cases}
\end{equation}
where $\mathcal{T}_a$ and $\mathcal{L}^{V/S}$ are functions of $y^a$. Under $h_{\mu\nu}\to h_{\mu\nu}+\bar{\nabla}_\mu\xi_\nu+\bar{\nabla}_\nu\xi_\mu$, the coefficients transform as
\begin{equation}
\begin{cases}
    \delta f_a^V = r D_a\!\left(\frac{\mathcal{L}^V}{r}\right) , \\
    \delta H^V = -\frac{k_V}{r}\,\mathcal{L}^V \,,
\end{cases}
\qquad \qquad
\begin{cases}
    \delta f_{ab} = D_a\mathcal{T}_b + D_b\mathcal{T}_a \,, \\
    \delta f_a^S = r D_a\!\left(\frac{\mathcal{L}^S}{r}\right) - \frac{k_S}{r}\,\mathcal{T}_a \,, \\
    \delta H^S = -\frac{k_S}{r}\,\mathcal{L}^S \,, \\
    \delta\gamma = \frac{k_S}{2r}\,\mathcal{L}^S + \frac{D^ar}{r}\,\mathcal{T}_a \,.
\end{cases}
\label{eq: KI gauge}
\end{equation}
It follows that
\begin{equation}
\begin{cases}
    F_a \equiv f_a^V + \frac{r}{k_V}\,D_aH^V \,, \\
    F \equiv \gamma + \frac{1}{2}\,H^S + \frac{D^ar}{r}\,X_a \,, \\
    F_{ab} \equiv f_{ab} + D_aX_b + D_bX_a \,,
\end{cases}
\qquad \qquad X_a \equiv \frac{r}{k_S}\left(f_a^S+\frac{r}{k_S}\,D_aH^S\right) ,
\label{eq: KI invariants}
\end{equation}
are gauge invariant, since $\delta X_a=-\mathcal{T}_a$. By virtue of the linearised Einstein equations, they can be expressed in terms of the master fields $\Phi^{(V/S)}$ as
\begin{equation}
\begin{cases}
    F^a = \frac{1}{r}\,\epsilon^{ab}D_b\Phi^{(V)} \,, \\
    F = \frac{1}{4}\,\Box_2\Phi^{(S)} \,, \\
    F_{ab} = \left(D_aD_b-\frac{1}{2}\,g_{ab}\Box_2\right)\Phi^{(S)} \,,
\end{cases}
\label{eq: KI master}
\end{equation}
where $\Phi^{(V/S)}$ satisfy the master equation \eqref{eq: master equation}. These are $r$ times the master fields of \cite{Anninos:2023epi}. The expression for $F$ agrees with the one given there upon using \eqref{eq: master equation}, together with $D_aD_br=0$ and $D_arD^ar=1$, which hold since $r$ is a Minkowski coordinate on the orbit space.

\textbf{Gauge choice.} We work in the gauge
\begin{equation}
    f^S_a = H^S = H^V = 0 \,,
    \label{eq: KI gauge choice}
\end{equation}
which is reached from a generic perturbation with $\mathcal{L}^S=\tfrac{r}{k_S}H^S$, $\mathcal{T}_a=X_a$ and $\mathcal{L}^V=\tfrac{r}{k_V}H^V$ in \eqref{eq: KI gauge}. It fixes the gauge completely. Preserving $H^{V/S}=0$ requires $\mathcal{L}^{V/S}=0$, and then preserving $f^S_a=0$ requires $\mathcal{T}_a=0$. In this gauge $X_a=0$, and the non-vanishing components of \eqref{eq: KI decomposition} coincide with the gauge invariants \eqref{eq: KI invariants},
\begin{equation}
    h^{(V)}_{ai} = rF_a\,\mathbb{V}_i \,, \qquad h^{(S)}_{ab} = F_{ab}\,\mathbb{S} \,, \qquad h^{(S)}_{ij} = 2r^2F\,\tilde{g}_{ij}\,\mathbb{S} \,.
\end{equation}
The scalar perturbation thus has no mixed and no traceless angular components, and the vector perturbation has no angular components. To recover \eqref{eq: KI ansatz}, we evaluate \eqref{eq: KI master} in the coordinates $(\tau,\rho)$. The non-vanishing Christoffel symbols of $g_{ab}$ are $\Gamma^\tau_{\tau\rho}=1/\rho$ and $\Gamma^\rho_{\tau\tau}=\rho$, such that
\begin{equation}
\begin{split}
    &D_\tau D_\tau\Phi = \partial_\tau^2\Phi-\rho\,\partial_\rho\Phi \,, \qquad D_\tau D_\rho\Phi = \partial_\tau\partial_\rho\Phi-\frac{1}{\rho}\,\partial_\tau\Phi \,, \qquad D_\rho D_\rho\Phi = \partial_\rho^2\Phi \,, \\
    &\Box_2\Phi = -\frac{1}{\rho^2}\,\partial_\tau^2\Phi+\partial_\rho^2\Phi+\frac{1}{\rho}\,\partial_\rho\Phi \,,
\end{split}
\end{equation}
which give the scalar part of \eqref{eq: KI ansatz}. For the vector part, $F_a=\tfrac{1}{r}\,\epsilon_{ab}D^b\Phi^{(V)}$ with $\epsilon_{\rho\tau}=\rho$, $D^\tau\Phi=-\rho^{-2}\partial_\tau\Phi$ and $D^\rho\Phi=\partial_\rho\Phi$ gives $rF_\tau=-\rho\,\partial_\rho\Phi^{(V)}$ and $rF_\rho=-\rho^{-1}\partial_\tau\Phi^{(V)}$.

\textbf{Physical diffeomorphisms.} Undoing the gauge transformation, a generic perturbation with $l\geq2$ reads
\begin{equation}
    h_{\mu\nu} = h^{(S)}_{\mu\nu}+h^{(V)}_{\mu\nu}+\bar{\nabla}_\mu\zeta_\nu+\bar{\nabla}_\nu\zeta_\mu \,, \qquad \zeta_a = -X_a\,\mathbb{S} \,, \qquad \zeta_i = -\frac{r^2}{k_S}\,H^S\,\mathbb{S}_i-\frac{r^2}{k_V}\,H^V\,\mathbb{V}_i \,,
\end{equation}
with $h^{(S/V)}_{\mu\nu}$ given by \eqref{eq: KI ansatz}, and $\zeta^\mu$ uniquely fixed by $h_{\mu\nu}$. Its normal component at $\Gamma$, $\zeta_\perp=-\sigma X_\rho\,\mathbb{S}$, does not vanish in general, so \eqref{eq: KI gauge choice} cannot always be reached with allowed diffeomorphisms. This is the origin of $h^{(\text{diff})}_{\mu\nu}$ in \eqref{eq: total pert}, which does not contribute to the gauge invariants. Since the vector part of $\zeta^\mu$ is tangent to $\Gamma$, physical diffeomorphisms only arise in the scalar sector.

\section{Late-time asymptotics} \label{app: late}

In this appendix, we derive the late-time behaviour \eqref{eq: late psi}. Consider the operator
\begin{equation}
    \mathcal{D}^{(l)} \equiv \left(l\tanh\tau-\partial_\tau\right)\left((l-1)\tanh\tau-\partial_\tau\right)\cdots\left(\tanh\tau-\partial_\tau\right) \,,
\end{equation}
which appears in \eqref{eq: PT recursion}. It satisfies the intertwining relation
\begin{equation}
    \left(\partial_\tau^2+V\right)\mathcal{D}^{(l)} = \mathcal{D}^{(l)}\,\partial_\tau^2 \,,
\end{equation}
such that $\psi=\mathcal{D}^{(l)}\chi$ solves \eqref{eq: PT} whenever $\chi$ solves the free wave equation $\partial_\tau^2\chi=\partial_x^2\chi$. In particular, $u_k=\mathcal{D}^{(l)}e^{ik\tau}$. In the absence of discrete modes, see section~\ref{sec: spectral}, every finite-energy solution in the inner region can be written in this form, with
\begin{equation}
    \chi(\tau,x) = F(x+\tau) + G(x-\tau) \,.
\end{equation}
Here $F$ describes waves moving towards $x\to-\infty$, and $G$ waves moving towards $\Gamma$.

The boundary conditions \eqref{eq: scalar BC} and \eqref{eq: vector BC} have constant coefficients in $x$, and therefore commute with $\mathcal{D}^{(l)}$. Writing them as $\mathcal{B}\psi|_{x=0}=0$, with $\mathcal{B}_S=\partial_x^2+\partial_x$ and $\mathcal{B}_V=\partial_x+1$, they become $\mathcal{D}^{(l)}\left(\mathcal{B}\chi\right)|_{x=0}=0$, that is,
\begin{equation}
    \mathcal{B}\chi\big|_{x=0} = \lambda(\tau) \,,
\end{equation}
where $\lambda$ belongs to the kernel of $\mathcal{D}^{(l)}$. This kernel is spanned by $l$ functions that are linear combinations of $e^{\pm j\tau}$, with $j=1,\ldots,l$. For instance, it is spanned by $\cosh\tau$ for $l=1$, and by $\cosh\tau$ and $\sinh2\tau$ for $l=2$. For initial data with compact support, $G(x-\tau)$ only contributes at $x=0$ for a finite interval of $\tau$. At late times, the boundary conditions thus become
\begin{equation}
    \text{Scalar:} \quad F''(s)+F'(s) = \lambda(s) \,, \qquad \text{Vector:} \quad F'(s)+F(s) = \lambda(s) \,,
\end{equation}
with $s=\tau$. The homogeneous solutions are $c_0+c_1e^{-s}$ for the scalar, and $c_1e^{-s}$ for the vector sector. The component of $\lambda$ proportional to $e^{-s}$ is resonant, and gives a contribution proportional to $s\,e^{-s}$. The components proportional to $e^{-js}$, with $j\geq2$, decay faster. Finally, the components proportional to $e^{js}$ give contributions to $F$ proportional to $e^{j(x+\tau)}$, which are mapped by $\mathcal{D}^{(l)}$ to $e^{jx}u_{-ij}(\tau)$. These are bounded, since $u_{-ij}\propto P_l^j(\tanh\tau)$ is a bound state of the P\"oschl--Teller potential, which decays as $\cosh^{-j}\tau$.

It remains to act with $\mathcal{D}^{(l)}$ at late times, where $\tanh\tau\to1$ up to corrections of order $e^{-2\tau}$. It maps a constant to $l!$ times the same constant, and $e^{-\tau}$ and $\tau e^{-\tau}$ to $(l+1)!$ times themselves, up to terms of order $e^{-\tau}$ in the latter case. We therefore find \eqref{eq: late psi}, with $u_\infty=l!\,c_0$. The rates are independent of $l$, since they are set by the boundary conditions at late times, where $V\to0$. These results agree with the numerical evolution in appendix~\ref{app: pulse}, see figure~\ref{fig: pulse boundary}.

\section{Numerical evolution of a compact pulse} \label{app: pulse}

In this appendix, we illustrate the results of sections~\ref{sec: spectral}--\ref{sec: late} by evolving compactly supported initial data numerically in the inner region. We consider the initial data
\begin{equation}
    \psi(0,x) = \mathfrak{b}(x+2) \,, \qquad \partial_\tau\psi(0,x) = \tfrac{1}{2}\,\mathfrak{b}(x+2) \,, \qquad \mathfrak{b}(s) = \begin{cases} \exp\left(4-\frac{4}{1-s^2}\right) \,, & |s|<1 \,, \\ 0 \,, & |s|\geq1 \,, \end{cases}
    \label{eq: pulse}
\end{equation}
supported in $\r e^{-3}<r<\r e^{-1}$ on the slice $t=0$. The data vanish in a neighbourhood of both $O$ and $\Gamma$, such that all corner compatibility conditions are satisfied. The qualitative behaviour is the same for any smooth initial data with compact support away from $O$ and $\Gamma$, in agreement with the energy estimate of section~\ref{sec: energy}. 

\begin{figure}[H]
\centering
\begin{subfigure}[b]{0.45\textwidth}
    \includegraphics[width=\textwidth]{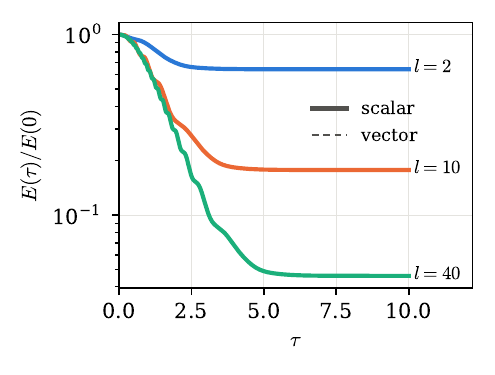}
    \caption{}
    \label{fig: pulse energy}
\end{subfigure}
\hfill
\begin{subfigure}[b]{0.45\textwidth}
    \includegraphics[width=\textwidth]{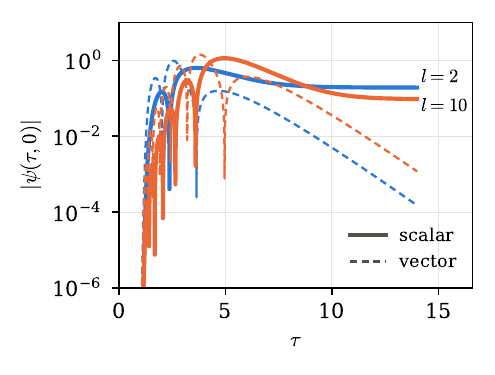}
    \caption{}
    \label{fig: pulse boundary}
\end{subfigure}
\\[0.5em]
\begin{subfigure}[b]{0.45\textwidth}
    \includegraphics[width=\textwidth]{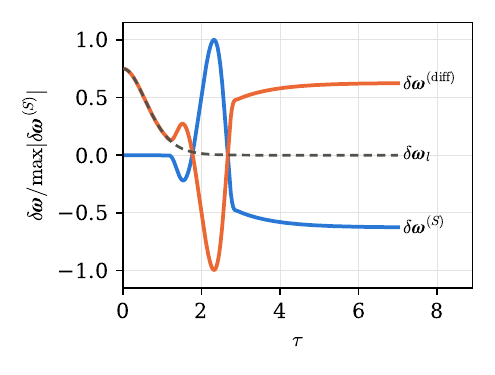}
    \caption{}
    \label{fig: pulse weyl}
\end{subfigure}
\hfill
\begin{subfigure}[b]{0.45\textwidth}
    \includegraphics[width=\textwidth]{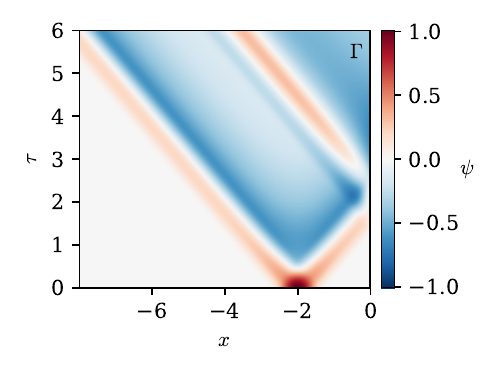}
    \caption{}
    \label{fig: pulse field}
\end{subfigure}
\caption{Evolution of the compact pulse \eqref{eq: pulse}. (a) Total energy \eqref{eq: energy} normalised to its initial value, for the scalar (solid) and vector (dashed) sectors with $l=2,10,40$. The two sectors nearly coincide. (b) $|\psi|$ at $\Gamma$ for the scalar (solid) and vector (dashed) sectors with $l=2,10$. The scalar one approaches a constant, while the vector one decays as $\tau e^{-\tau}$. (c) Weyl factors at $\Gamma$ for the scalar sector with $l=2$: the total Weyl factor $\delta\bomega_l$, with Weyl data \eqref{eq: pulse Weyl data} and $\alpha$ such that $\delta\bomega_2(0)=\tfrac{3}{4}\max|\delta\bomega^{(S)}|$, the part $\delta\bomega^{(S)}$ induced by the master field, and the normal displacement $\delta\bomega^{(\text{diff})}$, all normalised to $\max|\delta\bomega^{(S)}|$. (d) Scalar master field $\psi(\tau,x)$ for $l=2$, with $\Gamma$ at $x=0$. Minkowski time along $\Gamma$ is $t=\r\sinh\tau$, such that $\tau=6$ corresponds to $t\approx200\,\r$.}
\label{fig: pulse}
\end{figure}

Since the Weyl factor decouples from the bulk, the Weyl data do not affect the master fields. We set them to zero, except in figure~\ref{fig: pulse weyl}, where we take Weyl data on the decaying branch of \eqref{eq: Weyl l2},
\begin{equation}
    \delta\bomega_2=\frac{3\,\alpha}{\cosh^3\tau} \,.
    \label{eq: pulse Weyl data}
\end{equation}

\textbf{Numerical method.} We solve the master-field equation \eqref{eq: PT} in the inner region, $x\leq0$, with the initial data \eqref{eq: pulse} and the conformal boundary condition \eqref{eq: scalar BC} or \eqref{eq: vector BC} at $\Gamma$, located at $x=0$. We use the method of lines. Discretising $x$ on an equally spaced grid with second-order centred differences turns \eqref{eq: PT} into a system of ordinary differential equations for $\psi$ and $\partial_\tau\psi$ at the grid points, which we integrate with the classical fourth-order Runge--Kutta method. At $\Gamma$, the three-point stencil involves one auxiliary value outside the domain, which we eliminate using the centred discretisation of the boundary condition at every Runge--Kutta stage. The boundary value is then evolved like the interior ones and, in the scalar sector, obeys the discrete version of the dynamical boundary condition \eqref{eq: scalar BC dynamical}. We truncate the domain at $x=-X$, where we impose $\psi=0$. Since the data are supported in $-3<x<-1$ and propagate at unit speed, this artificial boundary does not affect the evolution up to $\tau=T$ for $X>T+3$; in practice we take $X\geq T+10$. The time step used is $\Delta\tau=0.4\,\min\big(\Delta x,1/\sqrt{l(l+1)+1}\big)$. %

\textbf{Accuracy checks.} Figure~\ref{fig: pulse} uses $\Delta x=0.01$, which we compare with $\Delta x=0.005$ and $\Delta x=0.0025$. For $l=40$, where the early-time oscillations are fastest, we additionally reduce $\Delta\tau$ by a factor of four. In both sectors, $\psi$ at $\Gamma$ and the total energy \eqref{eq: energy}, including the boundary energy \eqref{eq: boundary energy}, converge at second order in $\Delta x$, and the discrete energy decreases at every time step, as expected from \eqref{eq: energy derivative}. %

\textbf{Numerical results.} The results are shown in figure~\ref{fig: pulse}. The pulse splits into two parts. One part moves towards $x\to-\infty$ and leaves the inner region through the null cone $t=r$. The other is reflected at $\Gamma$, with $|\mathcal{R}_{S/V}|=1$, and then also moves towards $x\to-\infty$, see figure~\ref{fig: pulse field}. 

The total energy \eqref{eq: energy} decreases monotonically for both sectors and all values of $l$, see figure~\ref{fig: pulse energy}. At late times, $V\to0$ and \eqref{eq: PT} reduces to a free wave equation, such that the energy approaches a constant, carried by the waves that leave through the null cone. At $\Gamma$, $\psi$ approaches a constant in the scalar sector, while it decays as $\tau e^{-\tau}$ in the vector sector, in agreement with \eqref{eq: late psi}, see figure~\ref{fig: pulse boundary}. 

Figure~\ref{fig: pulse weyl} shows how $\Gamma$ accommodates the pulse. The Weyl factor $\delta\bomega_l$ evolves according to \eqref{eq: pulse Weyl data}, independently of the bulk. When the pulse reaches $\Gamma$, it induces the Weyl factor $\delta\bomega^{(S)}$ in \eqref{eq: Weyl S boundary}, and the normal displacement $\delta\bomega^{(\text{diff})}=\delta\bomega_l-\delta\bomega^{(S)}$ compensates for it, see \eqref{eq: xi perp solution}. At late times, $\delta\bomega^{(S)}$ approaches a constant, as discussed in section~\ref{sec: late}. In all these cases, no growth is observed.

\bibliography{bdryrefs}

\end{document}